\RequirePackage{fix-cm}
\documentclass[smallextended]{svjour3}       
\smartqed  
\usepackage{graphicx}
\journalname{Foundations of Physics}
\begin{document}
 
\title{Tests and problems of the standard model in Cosmology
}
\subtitle{}
 
 
\author{Mart\'\i n L\'opez-Corredoira
}
 
 
\institute{Instituto de Astrof\'\i sica de Canarias, E-38205 La Laguna, Tenerife, 
Spain \\
Departamento de Astrof\'\i sica, Universidad de La Laguna, E-38206 
La Laguna, Tenerife, Spain\\
              Tel.: +34-922-605264\\
              Fax: +34-922-605210\\
              \email{martinlc@iac.es}     
}
 
\date{Received: xxxx / Accepted: xxxx}

\maketitle
 
\begin{abstract}
The main foundations of the standard $\Lambda $CDM model of cosmology are that: 1) The redshifts 
of the galaxies are due to the expansion of the Universe plus  peculiar motions; 2) 
The cosmic microwave background radiation and its anisotropies derive from the high energy 
primordial Universe when matter and radiation became decoupled; 3) The abundance pattern
 of the light elements is  explained in terms of  primordial nucleosynthesis;  and 4)
 The formation and evolution of galaxies can be explained only in terms of gravitation 
within a inflation+dark matter+dark energy scenario. Numerous tests have been carried out
on these ideas and, although the standard model works pretty well in fitting many observations, 
there are also many data that present  apparent caveats to be understood with it. In this 
paper, I offer a review of these tests and problems, as well as some examples of alternative models.
\end{abstract}

\section{Introduction}
 
There is a dearth of discussion about possible wrong statements in the foundations of
standard cosmology (the ``Big Bang'' hypothesis in the present-day version of $\Lambda $CDM, i.e.
model which includes cold dark matter and the $\Lambda $ term in the energy, standing for
dark energy or quintessence). 
Most cosmologists are pretty sure that they have the correct theory, and that
they do not need to think about possible major errors in the basic notions
of their standard theory. Most works in cosmology
are dedicated to refining small details of the standard model and do not
worry about the foundations. There is still, however, a significant number of
results in isolated and disconnected papers that are usually
ignored by the leading cosmologists, and that are more challenging
and critical of the standard model. My intention here is to bring together
many of those heretical papers in order to help the more open-minded cosmologists 
to search the bibliography on tests and problems of the standard model.
 
In this paper, I will  critically review the most important assumptions
of the standard cosmological scenarios: 
 
\begin{itemize}
 
\item The redshifts of the galaxies are due to the expansion of 
the Universe.
 
\item The cosmic microwave background radiation and its anisotropies come from
the high energy primordial Universe.
 
\item The abundance pattern of the light elements is to be explained in
terms of the primordial nucleosynthesis.
 
\item The formation and evolution of galaxies can only be explained in
terms of gravitation in the cold dark matter theory of an expanding Universe.
 
\end{itemize}

Some observations will be discussed or rediscussed in order to show that
these facts were not strictly proven in some cases, but also, in other cases to show
the solidity of the standard theory against certain tests. 
 
There are many alternative theories.
However, the purpose of this review is not to analyze the different
theories, but to concentrate on the observational facts. 
In Ref.\ \cite{Lop14b} I have reviewed  theoretical ideas in cosmology that differ from the standard ``Big Bang''. It would be interesting
to extend this brief analysis to a wider literature survey, but that is a vast task well beyond the
scope of the present review. Another review paper would be necessary to compile these alternative ideas and complete 
the present review. Here, I focus mainly in the compilation of facts and their interpretations.
 
Cosmologists do not usually work within the framework of alternative cosmologies because they feel that
these are not at present as competitive as the standard model. Certainly, they are not so 
developed, and they are not so developed because cosmologists do not work on them. It is a 
vicious circle. The fact that most cosmologists do not pay alternative theories any attention and only 
dedicate their research time to the standard model is to a great extent due to the sociological 
phenomenon known as the ``snowball effect'' or ``groupthink''\cite{Lop14b}. This restricted view is unfair; therefore,
 I consider it appropriate to open the door here to further discussion of the results of the fundamental observations of cosmology.
 
It is not my purpose to defend
a particular theory against the standard cosmology. All theories have
their own problems.
Only the problems of the standard Big Bang theory are put forward here, but many more
affect the alternative models.
Not all the problems and all the papers are considered since the 
literature is too vast in this broad topic.
 
I have chosen to review  the general aspects of the 
foundations of cosmology as a whole instead of certain branches of it because I am interested in
expressing the caveats and open questions as a whole in order to extract
global conclusions on cosmology as a whole. 
It may also be that some of the caveats presented are no longer caveats, or
that some of the observational measurements are not correct. 
{\bf Warning:} {\it I  review only certain critical papers, and in a few cases
I discuss them, but I  take no responsibility for their contents.}
In most cases, I collect the information given by the different papers without any further
research; I am conscious in any case that most of the most critical papers need further analysis in
the problems they posit before reaching a firm conclusion on whether the standard model is correct or not. 
My own position is also neutral, I express no opinion on whether 
the standard cosmology is correct or not.

\section{Redshift and expansion}
 
\subsection{Is a static model theoretically impossible?}
 
A static model is usually rejected by most cosmologists.  
However, from a purely theoretical point of view, the representation
of the Cosmos as Euclidean and static is not excluded. Both expanding
and static space are possible for the description of the Universe.
 
Before Einstein and the rise of Riemannian and other non-Euclidean
geometries to the stage of physics, there were attempts to describe the
known Universe in terms of  Euclidean geometry, but with
the problem of justifying a stable equilibrium. Within a relativistic context,
Einstein\cite{Ein17} proposed a static model including a cosmological 
constant, his biggest blunder according to himself.
This model still has problems in guaranteeing stability, but it
might be solved somehow. Narlikar \& Arp\cite{Nar93} solve it
within some variation of the Hoyle--Narlikar conformal theory
of gravity, in which small perturbations of the flat Minkowski 
spacetime would lead to small oscillations about the line element 
rather than to a collapse. 
Boehmer et al.\cite{Boe07} analyze the stability of the Einstein static 
universe by considering homogeneous scalar perturbations in the context 
of $f(R)$ modified theories of gravity, and it is found 
that a stable Einstein cosmos with a positive cosmological constant 
is possible. Other authors solve the stability problem with the
variation of fundamental constants \cite{Van84,Tro87}.
Another idea by Van Flandern\cite{Van93} is that hypothetical gravitons 
responsible for the gravitational interaction
have a finite cross-sectional area, so that they can only travel a finite 
distance, however great, before colliding with another graviton. 
So the range of the force of gravity would necessarily be limited in this way
and collapse is avoided.
 
The very concept of space expansion has its own
problems\cite{Fra07,Bar08}.
Curved geometry (general relativity and its 
modifications) has no conservation of the energy-momentum of the gravity field. However, Minkowski space follows the
 conservation of energy-momentum of the gravitational field. One approach with a 
material tensor field in Minkowski space is given in Feynman's gravitation\cite{Fey95}, where 
the space is static but matter and fields can be expanding
in a static space. Also worth  mentioning is a model related to modern 
relativistic and quantum field theories of basic fundamental interactions 
(strong, weak, electromagnetic): the relativistic field gravity theory and fractal matter
distribution in static Minkowski space\cite{Bar08b}.

Olber's paradox for an infinite Universe also needs subtle
solutions, but extinction, absorption, and re-emission of light, fractal
distribution of density, and
the mechanism which itself produces the redshift of the galaxies might
have something to do with its solution.
These are old questions discussed in many classical books on
cosmology (e.g., \cite{Bon61}, ch. 3) and do not warrant further 
discussion here.

\subsection{Does redshift mean expansion?}
\label{.redshift}
 
Lema\^itre\cite{Lem27} in 1927 and later Hubble\cite{Hub29} in 1929 established the 
redshift ($z$)--apparent magnitude relation of the galaxies, which gave
an observational clue that the Universe is expanding.
Hubble was cautious in suggesting this interpretation, 
but the following generations of cosmologists became pretty sure that
the redshift of the galaxies following Hubble's law is a definitive proof of  expansion.
This was due mainly to the absence of a good theory that explains the possible phenomenological
fact of alternative proposals. General relativity provided an
explanation for the cosmological expansion, while alternative proposals were not supported
by any well-known orthodox theory. The expansion was preferred and the phenomenological
approaches which were not supported by present-day theory would
be doomed to be forgotten. This position would be right 
if our physics represented all the phenomena in the Universe, but
from a deductive-empiricist point of view we should deduce theories
from the observations, and not the opposite.
 
\subsection{Alternative redshift theories} 
 
There are other mechanisms that produce redshift\cite{Nar89,Bar94}
apart from space expansion or the Doppler effect. 
A bibliographical catalogue with hundreds of references before the '80s to 17 classes
of phenomenologies with untrivial redshifts is collected by Reboul\cite{Reb81}(Sect. 1),
the main idea being the ``tired light'' scenario.
 
A tired light scenario assumes that the photon loses energy owing
to some unknown  photon-matter process or photon-photon interaction, 
when it travels some distance: the distance is long if we consider all the 
intergalactic space between the object and the earth, or short,
for instance taking into consideration only the coronae enveloping
the object. Indeed, it is not so much a theory as a possible
phenomenological approach to explain the loss of the energy of the photons in a putative
static Universe that could be
explained by different theories. There are several hypothetical theories which
can produce this ``tired light'' effect. The idea of loss of energy of the photon
in the intergalactic medium was
first suggested in 1929 by Zwicky\cite{Zwi29} and
was defended by him for a long time. Nernst in 1937 had developed a model
which assumed that radiation was being absorbed by luminiferous ether.
As late as the mid-twentieth century, Zwicky\cite{Zwi57} maintained that
the hypothesis of tired light was viable.
But there are two problems\cite{Nar89}: 1) the $\phi $-bath 
smears out the coherence of the radiation from the source, and so all
images of distant objects would look blurred if intergalactic space
produced scattering, something that is incompatible with present-day 
observations\cite{Ste06}); 2) the scattering effect and the consequent loss of energy would be  frequency dependent, which is again  incompatible with what we observe in galaxies\cite{Rob72}\cite{Bar12}(S 4.1.1). Nonetheless, there are proposed solutions to these problems, as we will see in the following paragraphs.
 
Vigier\cite{Vig88} proposed a mechanism in which the vacuum behaves like 
a stochastic covariant superfluid aether whose excitations can interfere
with the propagation of particles or light waves through it in a
dissipative way. This avoids the two former difficulties of blurring
and frequency dependence. 
 
The photon-Raman interactions with atomic hydrogen\cite{Gal75,Mor06} 
also explains shifts that emulate the
Doppler effect with light-matter interaction that does not
blur the images. 
Or the redshift may be produced by a surface plasma that is undergoing 
rapid radial expansion giving rise to population inversion and laser action in some atomic
species\cite{Var79}.
 
The justification of the shift in photon frequency in a low
density plasma could also come from quantum effects derived from standard quantum
electrodynamics\cite{Lai97}. 
According to Paul Marmet and Grote Reber (a co-initiator of radio astronomy),
quantum mechanics indicates that a photon gives up a tiny amount of
energy as it collides with an electron, but its trajectory does not
change\cite{Ler91}(ap\-pen\-dix); this is called plasma redshift\cite{Bry04,Ash11,Wei14}. 
This mechanism also avoids
blurring and scattering. 
In another model\cite{Mam10} 
the photon is viewed as an electromagnetic wave whose electric 
field component causes oscillations in deep space free electrons, which then reradiate energy 
from the photon, causing a redshift. This new theoretical model is 
fundamentally different from Compton scattering, and therefore avoids any problems associated with Compton scattering, such as image blurring.
Potentially, this effect could explain the 
high redshifts of apparently nearby QSOs (Quasi Stellar Objects), 
since light travelling through the outer
atmosphere of the QSO could be redshifted before leaving it.
In order to explain galactic redshifts with long travel distances in the
scattering, the density in the intergalactic medium
should be 10$^4$ atoms/m$^3$, which is much higher than the density 
that is normally accepted ($\sim 10^{-1}$ atoms/m$^3$). 
However, the disagreement in the density of the intergalactic
medium is not necessarily a caveat in the hypothesis since our
knowledge of the intergalactic medium is very poor, and there is
also the possibility that intergalactic space is not so empty of
baryonic matter.

The dynamic multiple scattering theory is also very interesting
as a possible tired light mechanism.
Results in statistical optics\cite{Wol86,Roy00} have shown that the shift
in frequency of spectral lines is produced when light
passes through a turbulent (or inhomogeneous) medium, owing to multiple 
scattering effects. The new frequency of the line is proportional to the
old one, the redshift does not depend on the incident
frequency\cite{Roy00}. 
Redshift can also be explained in terms of non-linear optics, which assumes that the harmonic oscillator model of light has to be extended somewhat by an
extremely small anharmonic contribution\cite{Joo05}.
 
The redshift might be due to a tired-light gravitational 
interaction of wave packets with curved spacetime\cite{Cra06,Cra11a,Cra11c}. 
The result is a redshift that is proportional to distance and to the square root of the 
density. Curvature pressure results from the non-geodesic motion of 
charged particles in the gravitational interaction.

Apart from the standard
gravitational redshift derived from general relativity
\cite{Bon47,Bar81,Bar94b,Bar94}, with its own ca\-veats,
some theories are based on heterodox ideas about gravitation. Broberg \cite{Bro01},
for example, pointed out that a comparison between a quantizing scheme inherent in
the Schwarzschild metric and observed redshifts from quasars shows that the contracted
distances in the gravitational field are quantized in terms of the gravitational radii of the
gravitating objects responsible for the field, whilst non-contracted distances are not so
quantized. Indeed,
physicists have observed quantized states of matter under the influence of gravity: cold neutrons moving in a gravitational field do not move smoothly but jump from one height to another\cite{Nes02}.
In Broberg's view, this might be as important for modern quantum
physics as the Michelson-Morley Experiment was for the introduction of special relativity.
Quantization therefore appears in systems with significant relativistic contraction
(gravitational or Lorentz-contraction), such as de Broglie waves or, in the extreme case,
photon energy packets at the speed of light. In generalized terms, this would mean that
quantum physics has its origin in relativistically contracted fields, whereas the gravitational field can be scaled to particle dimensions, with a constant surface energy density as the
invariant parameter in the process. This might, for example, lead to the possibility of a
topological transformation from the gravitational force to the strong nuclear force.
 
Amitabha Ghosh\cite{Gho97} proposed a phenomenological model of dynamic
gravitational interaction between two objects that depends not only on their masses and the
distance between them, but also on the relative velocity and
acceleration between the two. These velocity- and acceleration-dependent interactions
are termed ``inertial induction''. The velocity-dependent force leads to a cosmic drag on all objects moving with respect to the mean rest-frame of a universe treated as infinite and quasi-static. This cosmic drag results in the cosmological redshift of light coming 
from distant galaxies without any universal expansion. This force model leads to the exact equivalence of gravitational and inertial masses and explains many not well understood 
observed celestial and astronomical phenomena.
 
``Self Creation Cosmology'' (SCC) \cite{Bar02a,Bar02b,Bar03a,Bar03b} 
 is an adaptation of the Brans Dicke theory in which 
the conservation requirement is relaxed to allow the scalar field to interact 
with matter. SCC can be thought of as general relativity + Mach's principle + local 
conservation of energy. In SCC energy is conserved but energy-momentum is not. 
Particle masses increase with gravitational potential energy and, as a 
consequence, cosmological redshift, is caused by a secular, exponential 
increase of particle masses. The universe is 
static and eternal in its Jordan frame and linearly expanding in its Einstein 
frame. Furthermore, as the scalar field adapts the cosmological equations, 
they require the universe to have an overall density of only one third of the critical density and 
yet be spatially flat. Also in the theory a ``time-slip'' exists between 
atomic ``clock'' time on one hand and gravitational ephemeris and cosmological 
time on the other, which would result in observed cosmic acceleration.
 
There is more to be said about gravity and/or interaction with gravitons.
In general relativity, dropping the restriction that a global time parameter exists, 
and assuming instead that the time scale depends on spatial distance, leads 
to static solutions, which exhibit no singularities, need no unobserved dark energy, and 
can explain the cosmological redshift without expansion\cite{Fis10}.
Additional redshift can also be explained using Weyl geometry-gravitation\cite{Bou78,Lun04,Kra08,Cas08}.
Interaction between gravitons and photons\cite{Van93,Iva04,Iva06} might be responsible for the redshift because it provides an energy loss mechanism that is not subject to the usual problem of making distant galaxy images fuzzy, and because it varies with $(1+z)^{-2}$
which is closer to the observed rate of decrease of surface brightness than any
Big Bang model. 
 
Instead of new ideas on gravitation, Roscoe's \cite{Ros06} aim was to 
shed new light on classical electrodynamics. After abstracting from 
certain well-known symmetries of classical
electrodynamics and showing that these symmetries are alone sufficient to recover the
classical theory exactly and unadorned, he found that the classical electromagnetic field
is then ``irreducibly'' associated with a massive vector field. 
This latter field can only be
interpreted as a classical representation of a massive photon. That is, it was demonstrated
that existing classical theory is ``already'' compatible with the notion of a massive photon.
He was then able to show that, given a certain natural assumption about how we should
calculate the trajectories of this massive photon, the ``quantized redshift phenomenology''
can be understood in a simple and natural way.
Other new ideas rested not on new physics or new reinterpretations of old theories, but
on effects predicted at present with standard physics.
Also, Mosquera Cuesta et al.\ \cite{Mos07} propose a non-cosmological redshift that is non-linear in terms of the electrodynamical  description of photon propagation through weak background 
intergalactic magnetic fields. 
Indeed, Maxwell may have been the first to consider the possibility of non-conservative photon behaviour. His original electromagnetic wave equation contained the 
energy damping term $\sigma _0\mu _0(\partial \phi/\partial t)$, where
$\sigma _0$ and $\mu _0$ represented the electrical conductivity and
magnetic permeabitity of background space\cite{Max91}(p.\ 431 of reprinted
version of 1954)\cite{Mon88}. In 1921, Nernst\cite{Ner21}(p. 40) put forth the idea in an
astronomical object by proposing that Olber's paradox might be resolved if
photons were assumed to undergo non-conservative energy damping during their
journey through intergalactic space.
 
Furthermore, just to give further examples from the huge literature on the topic,
there are explanations of non-cosmological/non-Doppler 
redshifts in terms of the local shrinkage of the quantum world\cite{Alf06}; or
the time variation of the speed of light\cite{Alf09}; or proposals of a 
quantum long time energy redshift\cite{Urb08}: the energy is smaller
owing to a quantum effect when the photon travels a very long time; 
chronometric cosmology\cite{Seg76,Seg95}; the variable mass hypothesis\cite{Hoy64,Nar77,Nar93}; 
time ac\-ce\-le\-ra\-tion\cite{Gar01}, etc.
 
All these proposed mechanisms show us that it is quite possible to
construct a cosmological scenario with non-expansion redshifts.
Nonetheless, all these theories are at present just speculations 
without direct experimental or observational support.
 
\subsection{Anomalous redshift in the laboratory and in the solar system}
 
There are not many experiments that try to measure anomalous  (neither Doppler nor
gravitational) redshifts in the laboratory.
One of these called my attention: Chen et al.\cite{Che09} reported the observation of line shift increase with plasma electron densities: 
roughly $\Delta \lambda ({\rm nm})\sim N_e(10^{18}{\rm cm}^{-3})$
for the HgI line with $\lambda =435.83$ nm. The dependence is also influenced by the temperature. This is justified by the authors 
by the difference in the atomic energy levels for higher electron density, leading
to a longer wavelength. The Hg atomic levels embedded in a density environment
are influenced by the free electron density. The electronic fields 
generated from free electrons compressed inside an atom screen the
Coulomb potential of the atomic nucleus. The nuclear forces on the
bound electrons are then diminished, while the repulsion of free to bound
electrons is enhanced, so that the energy levels outside the nucleus are 
raised and the intervals of excited energy level $7s^3S$ to $6p^3P_1^0$
are diminished. Previous studies also pointed out this kind of redshifts\cite{Ngu86}. My impression is that this emission redshift affects only to some lines and not the spectrum as a whole; in principle one would anticipate a dependence on wavelength.
Moreover, it is not shown here that this can explain the absorption of a photon and its emission with redshift (like cosmological redshift), as Ashmore\cite{Ash11} thinks.
 
The centre-to-limb variation of the
wavelengths of solar spectrum was discovered by Halm in 1907.
M\'erat et al.\cite{Mer74a,Mer74b} measured in 1974 redshifts in radio of background stars when
they are near the Sun, crossing the solar coronal layers, and it is
strong near the solar limb. General relativity predicts a deviation of light of $1.75^{\prime\prime}$.
However, very near the solar limb, the deviation is $2.03^{\prime\prime}$, 
whose excess over $1.75^{\prime\prime}$ might be linked with an excess of redshift.
Further research was carried out later by several defenders of anomalous
redshifts\cite{Mar89,Bry04}.
However, solar physicists consider the effect  
to be well explained in terms of standard physics and our knowledge of the convection in the Sun\cite{Dra82}. The properties of solar line asymmetries and wavelength
shifts can be traced back to the small-scale inhomogeneities in the
photosphere\cite{Dra82}: granulation (irregular polygonal shape of
1000-2000 km separated by narrow dark lanes).

\subsection{Observational tests for the expansion of the Universe}
\label{.testexp}
 
Nobody has ever directly observed a galaxy distance being increased with time. 
The motion of the putative expansion of
the Universe is so slow that it is unmeasurable. The variation in the redshift of a galaxy is also very slow and difficult to measure. Sandage\cite{San62} proposed in 1962
to measure the expected $\frac{dz}{dt}=(1+z)H_0-H(z)$ ($\sim $1 cm/s/yr), but it has not
hitherto been 
possible to meassure such a low acceleration. It would be possibly observable
with a 40 m telescope over a period of $\sim$20 yr using 4000 h of observing
time\cite{Lis08}. So, unfortunately, apart from the actual redshift of the galaxies, 
there are at present no direct proofs of the expansion. However, 
there are different indirect tests to verify whether the
Universe is expanding or static:
 
\begin{enumerate}
 
\item Microwave background temperature as a function of  redshift.
 
The Cosmic Microwave Background Radiation (CMBR) temperature 
can be detected indirectly at high redshift if suitable absorption lines can be found in high 
redshift objects. Hot Big Bang cosmology predicts that the temperature of the CMBR required to excite  these lines is higher than at $z=0$ by a factor $(1+z)$. For a static Universe, there
is no specific CMBR model, but one may say that the temperature would be constant and equal to 2.73 K for all redshifts if there were a local CMBR origin, or possibly $T_{\rm CMBR}(z)=2.73(1+z)$ 
K for a distant CMBR origin produced at a unique distance and with a redshift due to tired light.  
 
The CMBR temperature measured from the 
rotational excitation of some molecules as a function of redshift\cite{Mol02,Not11} has been quite successful in proving expansion: the results of 
Noterdaeme et al.\cite{Not11} with the exact expected dependence on
$T=T_0(1+z)$ are impressive. However, there are other results that disagree 
with this dependence\cite{Kre12,Sat13}. The discrepancy might be due to
a dependence on collisional excitation\cite{Mol02} 
or bias due to unresolved structure\cite{Sat13}.
 
Nonetheless, there is another piece of evidence of the relationship $T=T_0(1+z)$, which
comes from the Sunyaev--Zel'dovich effect\cite{Luz15}, 
the imprint of galaxy clusters on the CMBR, and this does not have the problems noted in the 
previous paragraph. Therefore, if we accept these results, 
at least one can say with almost total confidence that a local origin
of the CMBR in a static Universe is excluded, but the possibility of the origin of the CMBR at very high
$z$ within a static Universe is not excluded.
 
\item Time dilation test.
 
Clocks observed by us at high redshifts will appear to keep time at
a rate $(1+z)$ times slower when there is expansion. 
By using sources of known constant intrinsic periodicity,
we would expect  their light curves to be stretched in the
time axis by a factor  $(1+z)$.
 
Time dilation tests in Type Ia supernovae (SNIa) look set to become one of the most successful tests in favour of the expansion of the universe\cite{Gol01,Blo08}, but there are still problems in their interpretation. The fact that SNIa light curves are narrower when redder\cite{Nob08} is an inconvenience for a clean test free from selection effects. Other selection effects and the possible compatibility of 
the results with a wider range of cosmological models, including static ones, have also been pointed out\cite{Bry04b,Lea06,Ash12,Hol12}\cite{Cra11a}(Secc. 2)\cite{LaV12}(Secc. 7.8). 
 
Moreover, neither gamma-ray bursts (GRBs)\cite{Cra09} nor QSOs\cite{Haw10} present time dilation, which is puzzling.
However, 
with regard to this last fact, a variation in the quasar flux was found\cite{Dai12} in observations of more than 90 days with an average 
$|dF_V/dt|$ approximately proportional to $(1+z)^{-1}$, although they used a sample of only 
13 QSOs, without K-corrections. Further research is needed for the time dilation of
quasars to arrive at firm conclusions.
 
\item Cosmic chronometers.
 
Measurements of the Hubble parameter $\dot{z}$ using the variation of the age of elliptical galaxies with redshift as cosmic chronometers\cite{Mor12,Mor16} might be used to separate the different cosmological models.
For the standard cosmological model $-\frac{\dot{z}}{1+z}=H(z)=
H_0\sqrt{\Omega _m(1+z)^3+\Omega _\Lambda }$, whereas for a static model with tired light $-\frac{\dot{z}}{1+z}=H_0$, and for a static model with a linear extrapolation of Hubble's law with redshift $-\frac{\dot{z}}{1+z}=\frac{H_0}{1+z}$. The differences in the predictions are huge, and the actual data\cite{Mor12,Mor16} favour the standard model and exclude a static universe. Regrettably, however, the assumptions made for cosmic chronometers are incorrect. It is not true that elliptical galaxies have a single stellar population, they have at least two distinct stellar populations, a young one and an old one, and the use of the Balmer break to measure the age of the galaxy reflects an age difference between
both populations, not the maximum stellar age corresponding to the galaxy\cite{Lop16b}. This is a serious problem for the use of cosmic chronometers as they are applied
 nowadays. Therefore, so long we have no accurate methods to
determine the age of galaxies, this method is inapplicable.
 
\item Hubble diagram.
It has been known for  many decades that an apparent magnitude (taking into account K-corrections) vs.\ redshift diagram for elliptical galaxies in clusters better fits a static than an expanding Universe\cite{LaV86}.
This disagreement could, however, be solved by an increase of luminosity at higher redshift due to the evolution of galaxies.
 
For SNIa\cite{Kow08} or GRBs\cite{Wei10}, 
for which it is supposed there is no evolution, the standard model
works, provided that an ad hoc dark energy constant is included 
and we assume zero evolution and negligible extinction or selection 
effects (which are not universally accepted; \cite{Bal06,Pod08,Bog11}). 
Nonetheless, a static Universe may also fit those data\cite{Sor09,Ler09,Lop10a,Far10,Mar13}. 
By the way, there is anisotropy in the Hubble
diagram (values of $H_0$ and $\Omega _\Lambda $)
comparing data of SNIa with $z<0.2$ from
both Galactic hemispheres\cite{Sch07}.

\item The Tolman surface brightness test.
 
Hubble and Tolman\cite{Hub35} proposed the so-called Tolman test, based on the
measurement of surface brightness. A galaxy at redshift $z$ varies
in surface brightness proportionally to $(1+z)^{-n}$, with $n=4$ for expansion
and $n=1$ for the static case.

Lubin \& Sandage\cite{Lub01}, using the Tolman test up to $z=0.9$, claimed in 2001 to have definitive 
proof of the expansion of the Universe. 
However, their claim, rather than being a Tolman test, was that the evolution
of galaxies can explain the difference between the results of the Tolman
test and their preferred model, which includes expansion. 
Lerner\cite{Ler06} observed that Lubin \& Sandage used a very involved
evolutionary k-correction scheme, with many adjustable assumptions and
parameters to correct observed high-$z$ surface brightness. These evolutionary and K-corrections are subject to  uncertainty and cannot be used convincingly.
Crawford\cite{Cra11a} also pointed out that Lubin \& Sandage performed an erroneous analysis to exclude the static solution, mixing Big Bang and tired-light models.
 
Furthermore, other more recent Tolman tests\cite{Ler06,And06,Ler14,Cra11a}, some of them up to redshifts of $\sim$ 5 and with different wavelength filters so that no K-corrections are necessary, favour a static Universe without the need for galaxy evolution, but they cannot exclude the
standard scenario if evolution is allowed.
 
\item Angular size vs.\ redshift test. 
 
The angular size ($\theta $) of a galaxy with a given linear size\cite{Hoy59} is very different according to whether we assume the 
standard model with expansion or a static Universe. Tests have been made by 
several authors\cite{LaV86,Kap87,And06b,Lop10a},
and all of them, either in the radio, near infrared or visible, show, over a range of up to redshift 3,
a dependence $\theta \sim z^{-1}$, a static Euclidean effect over all scales. 
This result cannot be reconciled with the standard cosmological model
unless we assume a strong evolution of galactic radii that just coincidentally 
happens to compensate for
the difference\cite{Nab08}: galaxies with the same luminosity should be
six times smaller at $z=3.2$ than at $z=0$ \cite{Lop10a}.
It is usually further argued that elliptical and disc galaxies have different
evolutionary paths; that is right, but this difference was analysed and found to be compatible with
a null one if we take into account the error bars of selection effects\cite{Lop10a}.
Angular-size test for massive disc-like galaxies also obtain 
a good fit for static universes without evolution according to some authors\cite{LaV12}(Secc. 7.4)\cite{Ler15}.
Neither the hypothesis that galaxies which formed earlier have much
higher densities nor their luminosity evolution, merger ratio,
or massive outflows due to a quasar feedback mechanism
are enough to justify such a strong size evolution\cite{Lop10a}; also, 
the velocity dispersion would be much higher than observed\cite{Lop10a}.
A static Universe is fitted without any ad hoc element.
However, we must be cautious with this interpretation, because 
of the uncertainty in the galaxy size evolution.
 
There is yet another idea to be found in the literature about such a strong apparent evolution of galaxies:
at high redshift, owing to the Tolman factor proportional to $(1+z)^{-4}$, 
the surface brightness of the galaxies is much lower, and only very compact galaxies are detected
(with high surface brightness) for a given stellar mass/luminosity\cite{Dis12}.
This will give place to a certain ``apparent'' evolution in the average size of galaxies 
(smaller galaxies being observed at high redshifts), producing a downsizing illusion (dwarf galaxies apparently lift themselves above the sky at recent epochs).
 
So far, there are no standard rods serving as physical objects without evolution. 
Double radio sources were proposed as standard rods\cite{Val91} but there is also a strong
inverse correlation between their absolute brightness and linear size, and some degree of evolution is not totally excluded\cite{Val91,Nil93}. 
Ultra-compact radio sources were claimed in the past to be free of evolutionary effects, but they are now thought to show some signs of evolution as well\cite{Lop10a,Pas11}. The huge size evolution necessary to fit an angular size test with an expanding universe is not understood\cite{Lop10a} but still depends on our understanding of the galaxies rather than on pure cosmological approaches.
The ratio between gas mass fraction in clusters measured through X-ray and 
gas mass fraction (in the same clusters) measured through the Sunyaev--Zeldovich effect is proportional 
to the ratio of luminosity distance and angular distance, thus providing a method to derive indirectly the angular size (z), though this method assumes certain mass distributions 
in the cluster\cite{Hola12}. 
 
\item The Galaxy Number count magnitude test.
 
Differential galaxy number counts vs.\ galaxy K magnitude
are fitted by the standard cosmological model\cite{Tot01}, but also by the tired light
Universe\cite{LaV12}(Sect. 7.7). The same problem arises again: degeneracy between evolution 
and cosmological geometry.
 
\item UV surface brightness test.
 
Lerner\cite{Ler06} proposed a test of the evolution hypothesis 
that is also useful in the present case. There is a limit on the ultraviolet surface 
brightness (UV SB) of a galaxy because, when the surface density of hot bright stars and
thus supernovae increases, large amounts of dust are produced to absorb 
all the UV except that from a thin layer.
Further increase in the surface density of hot bright stars beyond a given
point just produces more dust and a thinner surface layer, not an
increase in UV SB. Based on this principle, there
should be a maximum UV(at-rest) SB independent
of redshift. This was analysed
at high redshift\cite{Ler06,Lop10a} and the result is that
the intrinsic UV SB magnitude of some galaxies would be prohibitively 
 lower(i.e.\ much brighter) than 18.5 mag$_{AB}$/arcsec$^2$ 
with the standard model.
For a static model, however, it would be
within the normally expected range. Lerner\cite{Ler06} also argues why  
alternative explanations (lower production of dust at high redshift, winds, 
or other scenarios) are inconsistent.
Nonetheless, Lerner's hypothesis of a maximum UV SB 
might be incorrect, so this should be further explored before
reaching definitive conclusions about this test.
 
\item Alcock--Paczy\'nski test.
 
Given a distribution of spherically symmetric objects with a radius along the line of sight
$s_\parallel=\Delta z\frac{d\,d_{\rm com}(z)}{dz}$
and a radius perpendicular to the line of sight
$s_\perp =\Delta \theta (1+z)^m d_{\rm ang}(z)$ ($m=1$ with expansion, $m=0$ for static),
the ratio $y\equiv \frac{\Delta z}{z\Delta \theta }\frac{s_\perp}{s_\parallel }$
depends on the cosmological
comoving distance ($d_{\rm com}(z)$) and the angular distance ($d_{\rm ang}(z)$),
and is independent of the evolution of galaxies, 
but it also depends on the redshift distortions produced by the peculiar velocities of  gravitational infall\cite{Lop14a}. 
 
I have measured $y(z)$ by means of the
analysis of the anisotropic correlation function of sour\-ces in several surveys\cite{Lop14a}, using a technique to disentangle the dynamic and geometric distortions, and also took other values available from the literature. From six different cosmological models (concordance $\Lambda $CDM, Einstein-de Sitter, open-Friedman Cosmology without dark energy, flat quasi-steady state cosmology, a static universe with a linear Hubble law, and a static universe with tired--light redshift), only two of them fitted the data of the Alcock \& Paczy\'nski's test: concordance $\Lambda $CDM and static universe with tired-light redshift. The rest were excluded at a $>95$\% confidence level. Analyses with further data using Baryonic Acoustic 
Oscillations (BAO)\cite{Mel15} improve the test and give us a more accurate constraint:
$\Lambda $CDM with standard values of parameters is excluded at 98\% C.L., whereas static with tired light fits the data, but also an expanding Universe with zero-active mass or wCDM with $\omega_{{\rm dark\ energy}}=-0.60^{+0.30}_{-0.27}$ and $\Omega_m= 0.45^{+0.21}_{-0.19}$ fits the data.

\end{enumerate}

\begin{table*}
\caption{Cosmological tests of the expansion.}
\begin{center}
\begin{tabular}{ccc}
\label{Tab:tests}
Test & Expansion & Static \\ \hline \hline
$T_{\rm CMBR}(z)$ & Good fit & $\begin{array}{l}
{\rm Tired\ light\ redshift\ of\ a\ CMBR}\\
{\rm coming\ from\ very\ high\ z.}
\end{array}$ \\ \hline
Time dilation & $\begin{array}{l}
{\rm Good\ fit\ for\ SNIa.} \\
{\rm Unexplained\ absence\ of\ time} \\
{\rm dilation\ for\ QSOs\ and\ GRBs.} 
\end{array}$ & $\begin{array}{l}
{\rm Selection\ effects,\ or\ ad\ hoc} \\
{\rm modification\ of\ the\ theory\ or} \\
{\rm the\ zero\ point\ calibration,\ or} \\
{\rm evolution\ of\ SNIa\ periods.}
\end{array}$ \\ \hline
Cosmic chronometer & $\begin{array}{l}
{\rm Good\ fit\ but\ by\ chance,} \\
{\rm measurements\ of\ differential\ age} \\
{\rm are\ incorrect.} 
\end{array}$ & $\begin{array}{l}
{\rm Bad\ fit\ but\ measurements\ of} \\
{\rm differential\ age\ are\ incorrect.} \\
\end{array}$ \\ \hline
Hubble diagram & $\begin{array}{l}
{\rm Good\ fit\ but\ requires\ the}\\
{\rm introduction\ of\ dark\ energy}\\
{\rm and/or\ evolution\ of\ galaxies.}
\end{array}$ & $\begin{array}{l}
{\rm Good\ fit\ for\ galaxies.\ Good} \\
{\rm fit\ for\ SNIa\ with\ some\ models.}
\end{array}$ \\ \hline
Tolman (SB) & Requires strong an SB evolution. &
Good fit \\ \hline
Angular size & $\begin{array}{l}
{\rm Requires\ too\ strong}\\
{\rm evolution\ of\ angular\ sizes.} 
\end{array}$ & Good fit \\ \hline
Galaxy counts & $\begin{array}{l}
{\rm Good\ fit\ for\ galaxies\ with}\\
{\rm evolution.} 
\end{array}$ & Good fit \\ \hline
UV SB limit & Anomalously high UV SB at
high z & Within the constraints \\ \hline
Alcock-Paczy\'nski & $\begin{array}{l}
{\rm Bad\ fit\ for\ the}\\
{\rm standard\ model\ but\ good\ fit}\\
{\rm with\ other\ wCDM\ models.} 
\end{array}$
& Good fit for tired light \\  \hline \hline
\end{tabular}
\end{center}
\end{table*}

Table \ref{Tab:tests} summarizes the analyses of this subsection. Apparently, there is no winner
yet. The first two tests favour expansion, whereas the 4--8th tests get a less ad hoc fit with the static solution, although this is insufficient to reject expansion. Most of the cosmological tests are entangled with the evolution of galaxies and/or other effects. Tolman or angular size tests need to assume a very strong evolution of galaxy sizes to fit the data with the standard cosmology, whereas the Alcock--Paczynski test is independent of the evolution of galaxies but
it does not show any preference for an expanding or static Universe yet.

\subsection{Anomalous redshifts}
 
Doubt might be cast upon the reality of the expansion as discussed above,
but even in the case that it is definitively proven, this does not mean that
all galaxies have a cosmological redshift, i.e.\ that all galaxies have a
redshift due to the expansion. There might be some exceptions, which are known
as ``anomalous redshift'' cases\cite{Nar89,Arp87,Arp03,Bur01,Bel02a,Bel02b,Bel06,Bel07,Lop06,Lop10b}; that is, a redshift produced by a 
mechanism different from the expansion of the Universe or the Doppler effect. 
 
There are plenty of statistical analyses\cite{Bur01,Chu84,Zhu95,Bur85,Bur96,Har00,Ben01,Gaz03,Nol05,Buk07,Bur09} showing an excess of 
high redshift sources near low redshift galaxies, positive and very 
significant cross-correlations between surveys of galaxies and QSOs, an 
excess of pairs of QSOs with very different redshifts, etc.
 
There are plenty of individual cases of galaxies with an excess of QSOs with high redshifts near the centre of nearby galaxies, mostly AGN (Active Galactic Nuclei). In some cases, the QSOs are only a few arcseconds away from the centre of galaxies with different redshifts. 
Examples are NGC 613, 2237+0305, and 3C 343.1. In some cases there are even filaments/bridges/arms apparently connecting 
objects with different redshift: in NGC 4319+Mrk 205, QSO1327-206, 
NGC 3067+3C232 (in the radio). The probability of chance projections of 
background/foreground objects within a short distance of a galaxy or onto 
the filament is as low as 10$^{-8}$, or even lower. 
The alignment of sources with different redshifts also suggests that they 
may have a common origin, and that the direction of alignment is the direction 
of ejection. This happens with some configurations of QSOs around NGC 4235, NGC 5985, GC 0248+430, etc. Other proofs presented in favour of the QSO/galaxy association with different redshift is that no absorption lines have been found in QSOs corresponding to foreground galaxies (e.g.\ PKS 0454+036, PHL 1226), or distortions in the morphology of isolated galaxies.
 
The non-cosmological redshift hypothesis also affects galaxies differently 
from QSOs. Cases such as NGC 7603, AM 2004-295, AM 2052-221, etc., present statistical anomalies also suggesting 
that the redshift of some galaxies different from that of QSOs might have 
non-cosmological causes. Not all supporters of the non-cosmological 
redshift agree with this idea; for instance, Arp claimed that galaxies might 
have non-cosmological redshift because they derive from an evolution of 
ejected QSOs, while Geoffrey Burbidge only defended the non-cosmological redshifts in QSOs.

In addition to the statistical anomalies, QSOs are still nowadays, half a century after their discovery, objects which are not completely understood in other aspects. There are many 
pending problems, inconsistencies, and caveats in the QSO research\cite{Lop11}. The standard paradigm model based on the existence of very massive black holes that are responsible for QSOS' huge luminosities, resulting from to their cosmological redshifts, leaves many facts without explanation. There are several observations that lack a clear explanation; for instance\cite{Lop11}, the absence of bright QSOs at low redshifts, a mysterious evolution not properly understood. There are alternative cosmological scenarios, such as the
decelaration of time\cite{Tag08}, that try to explain the
problem without referring to an evolution effect in which the position of QSOs in the Hubble diagram is explained by an improper estimation of luminosity.
 
Calculations of the probabilities of the anomalous redshift cases 
being normal background sources with cosmological redshifts and/or with
amplifications due to gravitational lensing\cite{Lop10b} indicate that
some of the examples of apparent associations of QSOs and galaxies with 
different redshifts may be just fortuitous cases in which background 
objects are close to the main galaxy although the statistical mean 
correlations remain to be explained, and some lone objects have a very low 
probability of being a projection of background objects. Nevertheless, 
these very low probabilities (down to $10^{-8}$ or even lower, assuming 
correct calculations) are not extremely low and, if the anomaly is real, one 
wonders why we do not find very clearly anomalous cases with probabilities as 
low as $10^{-20}$. Gravitational lensing seems not to be a general solution yet,
although it explains some of the anomalies\cite{Scr05}, and the requirement that the probabilities
be calculated a posteriori is not in general an appropriate answer for 
avoiding or forgetting the problem. 
 
There are two possibilities: either all cases of associations are lucky 
coincidences with a higher probability than expected for some still unknown 
reason, or there are at least some few cases of non-cosmological redshifts.  The debate has lasted a very long time, around 
50 years, and it is high time to consider making a last-ditch effort to finish 
with the problem. Every time the problem is mentioned, supporters of the standard view just smile or talk about ``a posteriori'' calculations, manipulations of data, crackpot ideas, without even reading any paper on the theme. The Arp--Burbidge hypothesis has become a topic in which everybody has an 
opinion without having read the papers or knowing the details of the problem, 
because some leading cosmologists have said it is bogus. 
On the other hand, the main supporters of the hypothesis 
of non-cosmological redshifts produce analyses of cases 
in favour of their ideas without too much care, pictures without rigorous 
statistical calculations in many cases, or with wrong identifications, 
underestimated probabilities, biases, use of incomplete surveys for 
statistics, etc., in many other cases\cite{Lop10b}. There are, however, many papers in which no objections are found in the arguments and they present quite 
controversial objects, but owing to the bad reputation of the topic, the 
community simply ignores them.

\section{Microwave Background Radiation}
\label{.CMBR}
 
 The 
CMBR has been interpreted as the relict radiation of an early stage of the Universe. Its black-body spectrum
 of 2.7 K reveals a very small dependence on sky position, of the order
of $\Delta T/T\sim 10^{-5}$ excluding the dipole due to the motion of the Earth.
Measurements of the anisotropies carried out by several teams of researchers
over the last three decades have been claimed to provide information 
on the structural formation of the
Universe, inflation in its early stages, quantum gravity, 
topological defects if any (strings, etc.), dark matter
type and abundance, dark energy or quintessence, the geometry
and dynamics of the Universe, the thermal history of the 
Universe at the recombination epoch, etc. 
Moreover, the CMBR has also become a source of accurate values for the parameters of this standard model of cosmology, which is usually called ``precision Cosmology''\cite{Pri05}. The foundations of cosmology are thought to be definitively established, and now a quantitative science is pursued for which the fitting of the power spectrum of the distribution of CMBR anisotropies and a few other data, and which gives us the numerical values of the parameters in the equations governing the entire past, present, and future of our Universe.

Gamow\cite{Gam53} and Alpher \& Herman\cite{Alp49} in the forties and fifties predicted an early stage of
the Universe in the Big Bang model that would produce a relic radiation
from this fireball that could be observed as a background.
However, Gamow and his coworkers were of the opinion that the
detection was completely unfeasible\cite{Nov01}.
The first published recognition of the relic radiation as a
detectable microwave phenomenon appeared in 1964\cite{Dor64}. 
None of the predictions of the background temperature based on the 
Big Bang, which ranged between 5 K and 50 K, 
matched observations\cite{van93b}, the worst being 
Gamow's upward-revised estimate of 50 K made in 1961. 
In 1965, the year of the discovery, a temperature of 30 K was calculated
for the amount of helium production observed\cite{Dic65}.
As the energy is proportional to $T^4$,
the energy observed is several thousand times less than predicted energy,
but it was predicted correctly that
it has a perfect blackbody spectrum\cite{Mat94}.
Although the discovery is attributed to Penzias \& Wilson, who won
the Nobel Prize because of it, the radiation was indeed previously 
discovered, although not interpreted in terms of a cosmological radiation:
in 1957, Shmaonov\cite{Shm57} was measuring radio waves coming from space at a wavelength
of 3.2 cm and obtained the conclusion that the absolute effective temperature
of radiation background appears to be 4$\pm $3 K, regardless of the 
direction of the sky. It is also possible that a team of japanese
radioastronomers measured this radiation at the beginning of 1950s\cite{Nov01}, and indirectly it was found by MacKellar in 1941 as the necessary radiation to excite rotating cyan molecules\cite{Nov01}; 
Herzberg also mentions in a book in 1950 \cite{Her50} that there is a strange excitation of molecular spectra, as if a 2.3 K radiation existed.
Anyway, the radiation is there and the question is the
origin of this radiation not the discoverer. 
Is the high energy primordial Universe the only possible scenario?
 
\subsection{Alternative explanations for the temperature of 2.7 K}
 
There were predictions of CMBR temperature with origin different from the 
Big Bang \cite{Ass95} by Guillaume, Eddington, Regener, Nernst, McKellar, Herzberg, Finlay-Freundlich, and Max Born.
Charles-Edouard Guillaume (Nobel laureate in Physics 1920) 
predicted in his article entitled ``Les Rayons X'' (``X-rays'', 1896) 
that the radiation of stars alone would maintain a background temperature 
of 6.1 K\cite{Mey03}. The expression ``the temperature of space'' is the title of 
chapter 13 of ``Internal constitution of the stars''\cite{Mey03,Edd26} 
by Eddington. He calculated the minimum temperature any body in space would cool 
to, given that it is immersed in the radiation of distant starlight. 
With no adjustable parameters, would be 3 K, essentially the same 
as the observed background (CMBR) temperature.
Other early predictions\cite{Mey03}, given by Regener\cite{Reg33} in 1933 
or Nernst\cite{Ner37} in 1937, gave a temperature of 2.8 K 
for a black body which absorbed the energy of the cosmic rays arriving on earth.
It was countered that Eddington's argument for the 
``temperature of space'' applies at most to our Galaxy. But Eddington's reasoning 
applies also to the temperature of intergalactic space, for which a minimum 
is set by the radiation of galaxy and QSO light. The original calculations 
half-a-century ago showed this limit probably fell in the range 
1--6 K\cite{Fin54}. And that was before QSOs were discovered 
and before we knew the modern space density of galaxies.
In this way, the existence of a microwave background in a tired light scenario\cite{Fin54,Bor54} was also deduced.
But in a tired light model in a static universe the photons 
suffer a redshift that is proportional to the
distance travelled, and in the absence of absorption
or emission the photon number density remains constant, we would not see a
blackbody background. The universe cannot
have an optical depth large enough to preserve a thermal background spectrum in
a tired light model\cite{Pee98} because we could not observe radio galaxies
at $z\sim 3$ with the necessary optical depth. Therefore, it seems that this
solution does not work, at least when the intergalactic medium instead of the
shell of the galaxy is responsible for the tired light.
 
In the fifties, it was pointed out\cite{Bon55,Bur58} that if the observed
abundance of He comes from hydrogen fusion in stars, there must have
been a phase in the history of the Universe when the radiation density
was much higher than the energy density of starlight today. If the
average density of the visible matter in the Universe is $\rho \sim
3\times 10^{-31}$ g/cm$^3$ and the observed He/H ratio by mass in it is 0.244 
(see \S \ref{.nucleo}), then the energy which must have been released in producing
He is $4.39\times 10^{-13}$ erg/cm$^3$. If this energy is thermalized, the black
body temperature turns out to be $T=2.76$ K, very close to the observed
temperature for the CMBR. Hence, there is a likely explanation
of the energy of the microwave radiation in terms of straightforward astrophysics
involving hydrogen fusion in stars\cite{Bur97b}.
Hoyle et al.\cite{Hoy68} also pointed out the suspect coincidence
between the microwave background temperature and that of hydrogen
in condensation on grains. They postulate that galaxy and star formation
proceed very readily when hydrogen is condensed on grains but not when
it is gaseous, and the universal microwave background temperature is that
associated with galaxy formation. 
The mechanism of thermalization in any of these cases
is perhaps the hardest problem.
 
\subsection{Alternative origin of CMBR}
 
Dust emission differs substantially from that of a pure blackbody.
Moreover, dust grains cannot be the source of the blackbody microwave radiation 
because there are not enough of them to be opaque, as needed to produce 
a blackbody spectrum. A solution for the blackbody emission shape
might be unusual properties of the dust particles: carbon needles, multiple explosions or big bangs, energy release of massive stars during the formation of galaxies, and so forth.
The quasi-steady state model\cite{Hoy93,Hoy94} argues that there is a distribution of
whiskers with size around 1 mm long and 10$^{-6}$ cm in diameter
with average $\sim 10^{-35}$ g/cm$^3$ providing optical depth 
$\tau \sim 7$ up to redshift $\sim 4$. 
However, the presence of a huge dust density to make the Universe opaque 
is forbidden by the observed transparency up to $z\sim 4$ or 5.
A solution might be an infinite universe.
An opaque Universe is required only in a finite 
space, an infinite universe can achieve thermodynamic equilibrium 
even if transparent out to very large distances. 
Somewhat similar to the proposal of whiskers is the proposal of the thermalization of ``cosmoids'', cosmic meteoroids which are also observed in the solar system\cite{Sob01},
or the proposal of emission by millimetre black holes\cite{Alf10}.
 
Another possible explanation is the
existence of an aether\cite{Clu80,Sor08}, i.e.\ a material vacuum, whose emission gives the
microwave background emission itself. This aether would be an incompressible
fluid according to Lorentz's theory and its existence is 
in opposition to Einstein's special relativity. 
Lorentz's\cite{Lor04} invariant laws of mechanics
in a flat space-time reproduce the standard observational tests of general
relativity provided that rest mass and light speed are not constant,
so the final choice between Einstein's and Lorentz's theories cannot
yet be regarded as settled according to Clube\cite{Clu80}.
 
The CMBR may be produced by the scattering of photons with electrons in a plasma of temperature
2 million K \cite{Bry04}.
Lerner\cite{Ler88,Ler95} proposes that electrons in intergalactic magnetic fields
emit and absorb microwave radiation. There is no relation between the
direction in which the radiation was moving when it was absorbed and its direction
on re-emission, so the microwaves would be scattered. After a few scatterings,
the radiation would be smoothed out. Magnetic fields much
stronger than the average field between galaxies would be needed;
perhaps the jets emitted from galactic nuclei would provide it.
The background radiation would be 
distorted by this intergalactic absorption against isotropy observations, so 
the radiation must instead come from the intergalactic medium itself in
equilibrium. This prediction agrees with the fact that the number of
radio sources increases much more slowly than the number of optical sources 
with distance, presumably owing to this absorption of radio waves in the 
intergalactic medium.
Observational evidence was also presented that something in the intergalactic medium
is absorbing radio and microwaves because farther radio sources with a given constant
infrared emission are fainter in the radio\cite{Ler90,Ler93}.  
However, there are some sources which are quite
bright in radio at intermediate redshift: 
Cygnus A ($z=0.056$) and Abell 2218 ($z=0.174$).
There is even a constant FIR/radio emission up to $z=1.5$ \cite{Gar02,Mao11} and sources 
are observed at $z=4.4$, so unless we have
a problem of anomalous redshift in all these cases, which seems unlikely,
Lerner's ideas do not work.
 
Related to electrons, there are other discussions of alternative explanations of CMBR
in the literature too. For instance, the radiation of electrons orbiting protons in atoms\cite{Shp02}. It is proposed that the atoms of hydrogen emit energy 
while in equilibrium owing to the orbiting of the electron around protons 
(which goes against the principles of quantum mechanics), thereby generating the background 
spectrum of 2.73 K. The atoms emit and absorb this radiation creating a 
background field with a blackbody shape.
 
Krishan\cite{Kri09} points out that, in addition to Thomson scattering, the absorption due 
to the electron--electron, electron--ion and the electron--atom collisions in 
a partially ionized cosmic plasma would also contribute to the optical depth 
of the cosmic microwave background (CMB). The absorption depth depends on the 
plasma temperature and the frequency of the CMB radiation. The absorption effects 
are prominent at the low frequency part of the CMB spectrum. These effects, 
when included in the interpretation of the CMB spectrum, may require a 
revised view of the ionization of the universe.
 
In ``Curvature Cosmology''\cite{Cra11b} the CMBR comes from the curvature-redshift process acting
on the high-energy electrons and ions in the cosmic plasma. The energy loss which gives way
to the spectrum of photons of the CMBR occurs when an electron that has been excited by the 
passage through curved spacetime interacts with a photon
or charged particle and loses its excitation energy.
 
The origin must be very local in order to preserve the blackbody shape, or be non-redshifted, which seems problematic.
Some proposals have even placed the origin of the CMBR within our own Galaxy, in the local bubble within 100 pc,\cite{Pec15}, positing that the microwave background radiation stems in very large part in re-radiation of thermalized Galactic starlight.
But the high level of isotropy, or the Sunyaev--Zel'dovich imprint of clusters of galaxies in the CMBR \cite{Agh11} or any possible correlation of the CMBR with  
galaxies would be puzzling in terms of such re-radiation.\footnote{This has indeed not been found yet, since the measurements of the cross-correlation of CMBR maps with galaxy surveys are not significant; see Ref. \cite{Lop10c} and references therein.}
 
Navia et al.\cite{Nav07} find that there is an isotropic distribution
of cosmic rays with energy $>6\times 10^{19}$ eV, so they should come from 
distances greater than 50 Mpc, and the cosmological interpretation
of the CMBR does not allow these cosmic rays to travel distances
$\ge 50$ Mpc \cite{Gre66,Zap66}.
However, Abraham et al.\cite{Abr07} claim that an isotropic distribution
of these cosmic rays is rejected with $>99$\% C.L. and that the
most probable sources are nearby AGN.
Kashti \& Waxman\cite{Kas08} also claim that the distribution
of these high energy cosmic rays is inconsistent with 
isotropy at $\sim $98\% CL, and consistent with a source distribution 
that traces LSS, with some preference to a source distribution 
that is biased with respect to the galaxy distribution.
 
In my opinion, at present there is no satisfactory alternative scenario
that does not have a problem in explaining the microwave background radiation, so
the standard scenario seems to be the best solution, although also with some discussions
about its plausibility
in some skeptical literature.\footnote{For instance,
the number of (CMBR) photons is much ($10^9$ times) higher than the number of cosmic baryons, thus indicating that cosmic evolution violates baryon number conservation; a heavy
baryon--antibaryon annihilation? Curiously, the CMBR photon density implies that the mean distance of photons is 0.2 cm, which, to the surprise of some, is just about identical to the maximum wavelength of the CMBR black body emission\cite{Fah09}.}
Nonetheless, one's mind should not be definitively closed against other options.
The hot primordial Big Bang is only a hypothesis that should not be
definitively 
considered as a solid theory just because of the Microwave Background Radiation.

\subsection{Microwave Background Radiation anisotropies}
\label{.anis}
 
Another fact is the existence of certain anisotropies in CMBR. There are still some authors
who doubt this fact, pointing out that 
there are spurious temperature anisotropies that are
comparable with the entire signal\cite{Li09,Rou10,Liu11}, or testing the null hypothesis that the Time-Ordered-Data (TOD) were consistent with no anisotropies when hourly calibration parameters were allowed to vary, i.e.\ that sky maps with no anisotropies outside the galactic band other than the dipole were a better fit to the uncalibrated TOD than 
those from the official analysis \cite{Cov09}. In my opinion, these analyses cannot be correct. The fact that the same anisotropies have been found by many different experiments and by different teams leaves  no doubt in my mind that the anisotropies are real.
 
The first predictions of CMBR anisotropies were wrong. One predicted 
$\Delta T/T$ to be one part in hundred or thousand \cite{Sac67}; however, this value could not fit the observations, which gave values hundreds of times smaller, so non-baryonic dark matter was introduced {\it ad hoc} to solve the question.

CMBR analyses in the last two decades have concentrated on anisotropies on small angular scales, smaller than the angular resolution of several degrees achieved by COBE-DMR in the '90s. The power spectrum of the a\-ni\-so\-tro\-pies of the experiments
BOOMERANG and MAXIMA-1 showed a first peak at the Legendre multipole 
$\ell \approx 200$,\cite{deB00,Han00} 
corresponding to angular scales of $\approx 1.2^\circ $, which was interpreted as a discovery of the previously predicted acoustic peaks in a flat Universe with 
$\Omega =\Omega _m+\Omega _\Lambda =1$, the greatest contribution coming from a dark energy component of $\Omega _\Lambda =0.7$ (see review in Ref. \cite{Hu02} and references therein for further comments). Indeed, there are two parts in the comparison 
of theoretical predictions and observations: a successful one, and a half-successful/half-failed one. The totally successful prediction was the qualitative shape of the power spectrum containing several peaks: For instance, Peebles \& Yu\cite{Pee70} had predicted a CMBR power spectrum with fluctuations, the acoustic peaks, from the hypothesis of primaeval adiabatic perturbations in an expanding universe. However, the position of the first peak was not at the expected position when first measured: already in the mid-'90s the position of the first peak was determined to be $\ell \approx 200$  with other experiments to measure small angular resolution anisotropies before BOOMERANG or MAXIMA-1 but with smaller sky coverage (Ref. \cite{Whi96} and references therein). De Bernardis et al.\cite{deB00} and Hanany et al.\cite{Han00} just added a refinement in the measurement of the position of the first peak but they were not its discoverers. White et al.\ in 1996 \cite{Whi96} realized that the preferred standard model of the time (an open Universe with 
$\Omega =\Omega _m\approx 0.2$ and without dark energy) did not fit the observations, so that they needed a larger $\Omega $. This was one of the elements, together with SNIa observations, and the age problem of the Universe, that would encourage cosmologists to include the new {\it ad hoc} element: dark energy. Between 1997 and 2000 this change of mentality in standard cosmology was produced, and then, in 2000, with the results of the new BOOMERANG and MAXIMA-1 experiments, cosmologists were proud to announce that new observations were giving exactly the results they expected. In any case, just paying attention to the first analyzes of the first peak in the mid-'90s, I would attribute a half-success to the prediction because, even though they failed to fit the observations with the preferred standard model of a curved open Universe at that time, the idea of a flat Universe was also presented as a possibility in the '90s, and indeed a possibility preferred by inflationary paradigms.
The position of the other peaks would also serve to constrain the cosmological models. The acoustic peaks on angular scales of 1 deg and 0.3 deg were predicted with the second peak nearly as high as the first one in $\ell (\ell +1)C_\ell/(2\pi )$ \cite{Bon87,Jor95}. Later, as data became available, its amplitude was reduced and the positions of the other peaks and their relative heights were also constrained from the model with a higher agreement with the data (e.g.\ Ref. \cite{Hu01}).
 
More recently, the analysis of CMBR polarization anisotropies (e.g.\ Ref. \cite{Lar11}) has also provided strong support for the standard cosmology. This indicates the way in which the CMBR is polarized owing to the scattering of free electrons. Photon diffusion into regions of different temperatures are possible only when the plasma becomes sufficiently optically thin; these diffused photons could then scatter only while there are still free electrons left. Since photons could not diffuse too far, polarization cannot vary much over very large angular scales.
 
Given this history of CMBR analyses, it is difficult to accept that 
our observations reproduce ``predictions''. Nonetheless, whichever comes first,  theory or  observation, the fact remains that we have now a cosmological model that is able to fit  the CMBR power spectrum $C_\ell$ quite accurately. 
It is usually thought that this may not happen by chance, given the apparent complexity of the power spectrum shape, whereas the six\footnote{Plus many other parameters which introduce second-order changes. And, even so, there is a degeneracy in the solutions with different values of $H_0$ and $\Omega _\Lambda $: CMBR data, and the large scale structure of galaxies could be reproduced without explicitly requesting the existence of dark energy\cite{Bla03} i.e.\  with $\Lambda =0$. This degeneracy is broken by adding cosmological information from other sources, for instance, from SNIa data. In order to fit the temperature--polarization cross power spectrum and the polarization--polarization power spectrum\cite{Lar11}, one would need an extra parameter (optical depth), so a total of at least seven free parameters are necessary.
Roughly speaking, the relationship between temperature-temperature and the polarization--temperature, or polarization--polarization power spectra is expected since they are different ways of seeing the same light with different filters.} parameter cosmological model is relatively simple\cite{Lar11,Pei05}, but this is not correct, as I will explain in the following paragraphs. 
 
Some critics of the standard model\cite{Dis07} claim that there is little value in the fact that the standard cosmological model might fit some observational cosmological data if the number of free parameters in the model were comparable to the number of independent parameters  characterizing the observations. For instance, Ptolemaic geocentric astronomy may fit the observations of the orbits of the planets but with too high a number of free parameters in the theory. 
The principle of Occam's razor tells us that a theory is better when the number of free parameters is low. Occam's razor can indeed be understood with a Bayesian analysis, in which it is tested how probable  a model  is with a given number of free parameters with respect to other
models with a higher number of free parameters\cite{Jef92,Ber92}. Otherwise, if a theory has a high number of free parameters, it loses credibility because it is always possible to create a ``false'' model to fit some data when the number of free parameters is comparable to the number of degrees of freedom in the data.
Philosophers of science, when talking about cosmology,  associate this approach with ``instrumentalism'' (e.g., Ref. \cite{Sol08}).
And saying that the power spectrum contains hundreds of independent parameters for a given resolution is not correct, because the different values of $C_\ell $ for each $\ell$ are not independent in the same sense that hundreds of observations of the position and velocity of a planet
do not indicate hundreds of independent parameters. Indeed, the information on the orbit of planet is reduced to only six Keplerian parameters.
 
For the reasons just given, understanding how much information is in the power spectrum is important for key questions in the discussion of the fundamentals of cosmology. 
There are indeed two main points which should be clarified:
\begin{enumerate}
\item Are the oscillations something atypical in a power spectrum? That is, should we consider the fact the power spectrum contains oscillations a successful prediction of the standard cosmological model that cannot be produced by any other means? 
\item How much information is contained in the power spectrum? That is, how many free parameters in a function are necessary to fit the CMBR power spectrum? Should we consider the fitting of the power spectrum with a model of six free parameters as a validation of the standard cosmological model? 
\end{enumerate}
 
 For the answer to the first question, we must bear in mind that
the presence of peaks in the power spectrum  is a rather normal characteristic expected from any fluid with  clouds of overdensities that emit/absorb radiation or interact gravitationally with the photons, and with a finite range of sizes and distances for those clouds. Apart from the standard cosmological model, other scenarios may also follow these conditions. The interpretation of ``acoustic'' peaks is just a particular case; peaks in the power spectrum may be generated in scenarios that have nothing to do with oscillations due to gravitational compression\cite{Lop13a,Lop13b}.
 
For the second question, we must note that a 
simple polynomial function with no physical interpretation with six free parameters is able more or less to reproduce the two-point correlation function (the Fourier transform of the
power spectrum), giving very good fit for at least the 
first two peaks of the power spectrum\cite{Lop13b}.
The fact the standard model is able to fit the CMBR anisotropies with a model with six free parameters is astonishing but it would be much more surprising if it could fit it without free parameters or with only one or two free parameters fitted to the power spectrum. Certainly, the same parameters which are used to fit the CMBR power spectrum are able to
fit other cosmological data, but more than six parameters are necessary, as well as additional
information such as initial conditions, conditions of stellar formation, galaxy formation, how dark matter is distributed in galaxies, etc. A global analysis of the cosmological models would require the examination of all of the available independent sources of cosmological
data (nucleosynthesis, supernovae, gravitational lensing, etc.) and to check whether they are comparable to the number of free parameters in the model.
 
In  cosmologies different from the standard model, there were also attempts to fit the CMBR power spectrum with models with  a similar number of free parameters.
Narlikar et al.\cite{Nar03,Nar07} fitted the $C_\ell $ with five--six parameters apart from the amplitude within a model that has nothing to do with the origin of the CMBR in the standard cosmology.
Power spectrum peaks at $\ell \approx 6$ to 10, 180 to 220, and 600 to 900 are shown to be respectively related in their Quasi-Steady State cosmology to curvature effects at the last minimum of the scale factor, clusters, and groups of galaxies. 
Previously, it had been calculated that clusters of cold clouds in the
halo produce anisotropies, which peak at $l\approx 50$, has a power 
proportional to $1/sin^2b$ and can represent around 5\% of the
total anisotropies at $b=30^\circ $ \cite{Wal03}.
For a  MOND (Modified Newtonian Dynamics) cosmology without cold dark matter, it is also possible with few parameters to fit the power spectrum\cite{McG04,Ang11}.
It is not clear, however, that these alternative scenarios can explain
the phase coherence needed to account for the clear peak/trough structure observed in
CMBR anisotropies, as predicted by an inflationary scenario in the standard model\cite{Dod03}.
 
\subsection{Non-gaussianity}
\label{.gauss}
 
Another relevant feature of CMBR anisotropies is the gaussianity of its fluctuations. 
Theories involving inflation generally predict a pattern of Gaussian noise, whereas theories based on symmetry breaking and the generation of defects have more distinctive signatures\cite{Cou94}. 
However, this should not be a major problem for alternative cosmologies since many different origins of the fluctuations are expected to be Gaussian. And even if there may be phenomena that generate deviation from normal Gaussian fluctuations, if many of them intervene, the result will be Gaussian anyway.
 
It is not totally clear yet whether the CMBR fluctuations are exactly Gaussian: a non-Gaussian distribution has been
claimed by several authors (e.g., Ref. \cite{Lop07}, \S 2.3 and references therein; 
Refs. \cite{Fer98,Pan98,Jeo07,Rae07,Ber07,McE08,Ros09}); however, other analyses\cite{Rub06,McE06} claim that only a few regions have such non-Gaussian anisotropies owing to contamination,  e.g.\ the Corona Borealis supercluster region, and most of the regions in the sky are Gaussian. The non-gaussianity may be associated with cold spots of unsubtracted foregrounds\cite{Liu05,Toj06}; even the lowest spherical harmonic modes, which should be the cleanest, in the map are significantly contaminated with foreground radiation\cite{Chi07}.

This might mean that analyses claiming 
non-gaussianity are wrong, or that the gaussianity is not significant enough, or that the
inflation model is incorrect, or that some contamination is present
in the maps, which are supposed to be clear of any contamination.
Among all these possibilities, maybe more than one applies, and I suspect
that one of these is the incorrect subtraction of the Galactic contamination,
although other factors may also be important too.

\subsection{Some doubts on the validity of the foreground
Galactic contribution subtraction from microwave
anisotropies}
\label{.Galcont}

Many authors  have studied the different components
of the Galactic microwave foreground radiation, but in my view these efforts are still insufficient to separate the components appropriately. The errors in the foreground 
emission subtraction are small, but they are not negligible.
At least, some doubts on the validity of the foreground Galactic subtraction from 
microwave anisotropies can be expressed\cite{Lop07} . 
 
The question of the Galactic foregrounds has been present for a long time now. However, an analysis of the papers analysing the problem
reveals that it has sometimes been underestimated, and that the problem of Galaxy subtraction has become more complex in recent years.
For instance, two decades ago it was thought\cite{Gut95,Dav96} that Tenerife data at
15 GHz were likely to be dominated by cosmological fluctuations, and
that the 33 GHz data were scarcely affected by the Galaxy ($\sim 4$ $\mu $K 
in scales of 5--15 degrees). Later analyses (WMAP, \cite{Ben03b}(Fig. 10)), 
however, have shown that for comparable angular
scales the Galaxy is dominant in the anisotropies at 22.8 GHz
(and the Galactic contamination at 15
GHz is not much smaller than at 22.8 GHz), and at 33 GHz the Galactic 
anisotropies are of the same order as those from the CMBR. 
Also, at small scales, less than a degree, the foreground Galactic emission
dominates\cite{Lei00}.  An
explanation for this is that it is mainly due to the existence
of a kind of emission unknown in the early '90s, which is correlated with the
dust that was not discovered previously; the synchrotron and 
or free--free emission might also have been
underestimated. Positive correlations between the 
microwave anisotropies, including the region around
15 GHz, and far-infrared maps, which trace Galactic dust, 
were found\cite{Lei00,Kog96,Fin04,Fer06}. 
In the Helix region
the emission at 31 GHz and 100$\mu $m are well
correlated\cite{Cas04}; the 100 $\mu $m-correlated 
radio emission, presumably due to dust, accounts for
at least 20\% of the 31 GHz emission in the Helix (so the total 
dust emission is higher than 20\% because
there is a non-correlated component too). 
The anomalous emission at
around 23 GHz of the Perseus molecular cloud (temperature $\sim 1$ mK)\cite{Wat05} 
is an order of magnitude larger than the
emission expected from synchrotron + free--free + thermal dust. There is also some correlation of 10, 15 GHz maps with $H_\alpha $
maps, but very low, so the free--free emission is detected at 
levels far lower than the dust correlation\cite{deO99,deO02}.
The most likely current explanation for this
emission correlated with dust around 15--50 GHz is spinning 
dust grain emission\cite{Dra98} and/or magnetic dipole emission from ferromagnetic 
grains\cite{Dra99}. 
More recently, a new foreground was discovered for low frequencies: ``microwave haze'' emission around the Galactic 
center\cite{Pie12}.
Whatever it is, it is now clear that the
15--40 GHz range is dominated by the Galaxy.
A question might arise as to whether the contamination in the
remaining frequencies (40--200 GHz)
is being correctly accounted for. Typical calculations for 
dust contamination claim that it should not be
predominant, and that its contribution can be subtracted accurately, 
but how sure can we be of this
statement? How accurate is the subtraction of the dust 
foreground signal?
 
A first difficulty in the subtraction of the dust component is to
know exactly how much emission there is in
each line of sight. First approximations came with extrapolations 
from the IRAS far-infrared and DIRBE
data (with the zodiacal light subtracted, as well the cosmic
infrared background) used to model the dust
thermal emission in microwaves. Templates were taken from these 
infrared maps and extrapolated in
amplitude by a common factor for all pixels. The problem is 
that, as said by Finkbeiner et al.\cite{Fin99},
{\it ``a template approach is often carelessly used to compare observations with expected contaminants, with the correlation amplitude indicating the level of contamination. (...) These templates ignore well-measured variation in dust temperature and variations in 
dust/gas ratio.''}
The growing contrast of colder clouds in the background of 
the diffuse interstellar medium will produce
much higher microwave anisotropies than the product of the 
template extrapolation\cite{Lop99b}. Neither is it a good
strategy to subtract a scaled IRAS template to remove the spinning dust in multifrequency data since it
produces large residual differences\cite{Lei00}.
There is a presence of residual foreground emission not traced by the templates\cite{Bot08}.
A better approximation to this dust emission in the 
microwave region came with the adoption of
an extrapolation with colour corrections in each pixel. 
This method assigns a different temperature to
each pixel. This colour correction, together with a
$\nu ^2$
emission emissivity\cite{Sch98}, gave
a much tighter agreement with FIRAS data. However, it was 
still inconsistent with the FIRAS data
below 800 GHz in amplitude\cite{Fin99}: a 14\% error 
in the mean amplitude at 500 GHz.
Indeed, no power-law emissivity function fits the FIRAS data
in the 200--2100 GHz region\cite{Fin99}. 
Furthermore, laboratory measurements suggest 
that the universality of $\nu ^2$
emissivity is
an oversimplification, with different 
species of grains having different emissivity laws\cite{Fin99}. 
A better approximation is an extrapolation
with two components ($\nu ^{1.7}$, $\langle T\rangle $=9.5 K and $\nu ^{2.7}$,
$\langle T\rangle $=16 K); each pixel has an assignation of two temperatures 
\cite{Fin99} with correction factors that are a function of 
these temperatures. This still fails in the predictions of
FIRAS from the IRAS--DIRBE extrapolation by 15\% in zones dominated 
by atomic gas\cite{Fin99}.
 
Finkbeiner et al.'s\cite{Fin99} approach in 1999, although
much better than a direct extrapolation of the template, was insufficient.
The problem is difficult to solve because it is
an extrapolation by a factor $\sim$15--60 in frequency (from 1250 GHz
[240 $\mu $m] or 3000 GHz [100 $\mu $m]) to 50--90 GHz). This is equivalent,
for instance, to the attempt to derive a map of stellar emission in
12$\mu $m as an extrapolation of the emission in the optical B-filter.
Frankly, when I see the IRAS-12$\mu $m map of point sources and
the Palomar plates in blue filters, I observe huge differences, 
and I do not know how we can extrapolate
the second map to obtain the first one. Each star has a different
colour, and stars which are very bright in blue may be very
faint at mid-infrared and vice verse. The same thing happens with the
diffuse + cloud emission: there are hot regions, cold regions, different kinds
of emitters (molecular gas, atomic gas) and we have to integrate all
this into each line of sight. The assumption that with only two temperatures
we can  extrapolate the average flux, and that a colour term can
correct the pixel-to-pixel differences is comparable to the assumption that 
a model with only two kinds of stars and the knowledge of $(B-V)$ for each star
we can extrapolate the star counts from optical to mid-infrared 
for the whole Galaxy.
It is also very common\cite{Mas01,Fer06} to calculate the Galactic dust contribution
in some microwave data by just making a cross-correlation between
these data and some far-infrared map of the sky. This is simply
wrong and not even valid for ascertaining the order of magnitude of
such contamination. 
Nowadays, things are carried out with much better data and with more
frequencies (for instance, with the PLANCK data\cite{Pla15}), but the philosophy of extrapolation and correlation is still similar, and the maps which are supposed to be free of
contamination still present signs of dust contamination\cite{Axe15}.
 
Therefore, any method of foreground subtraction which uses templates
will have serious credibility problems with regard to the goodness of the
subtraction. This applies not only to those methods that use templates directly with
coupling coefficients derived from cross-correlation
but also to MEM (maximum entropy method; \cite{Ben03b}), which uses
in the initial stage templates for the dominant foreground components
and also establishes some a priori conditions of their spectral behaviour.
Moreover, any calculation of the limits of such contamination based on
cross-correlations will not be totally accurate.
 
The presence of different dust-emitting components causes 
the spectral indices of the foregrounds to vary with position.
Spectral index changes 
from $-$5.1 to $-$2.1, depending
on the region have been measured\cite{Kis03}, and also changes with wavelength; they
point out the existence of dust at different temperatures,
in particular of a cold, extended component.
A catalogue of Galactic cold clumps derived with PLANCK 
gives spectral index between 1.4 and 1.8 \cite{Ade11}.
Even small spectral index variations as small as $\Delta \alpha \sim 0.1$
can have a substantial impact on how channels should be combined and on the
attainable accuracy\cite{Teg98}, so the errors of the subtraction 
assuming certain power spectrum for the dust are serious.
 
Another technique that does not use a templates is
ILC (internal linear combination)\cite{Ben03b}. It assumes nothing about
the particular frequency dependencies or morphologies of the foregrounds and
tries to minimize the variance in different regions of the sky
with the combination of the available frequencies.
There is a degeneracy of solutions, an infinite number of maps can
be generated, and there is no way to test
whether the maximum likelihood solution is the correct one\cite{Rob07}.
Those that have applied this method\cite{Ben03b} warn against its use for cosmological
analysis; it is not effective in removing all residual foregrounds\cite{Eri04b}.
The ILC method performs quite badly, especially for dust\cite{Eri04b},
in part because of the variability of the spectral indices. 
Therefore, ILC maps are not clean enough to allow cosmological conclusions to be arrived at \cite{Eri04b}: {\it `[The] ILC map, which by eye looks almost 
free of foreground residuals, has been extensively used for scientific
purposes---despite the fact that there are strong (and difficult to
quantify) residual foregrounds present in the map. (...) the ILC map
is indeed highly contaminated by residual foregrounds, and in particular,
that the low-$\ell$ components, which have received the most attention so
far, are highly unstable under the ILC cleaning operation'} (Eriksen
et al.\cite{Eri05}). 
ILC provides satisfactory results only under rather 
restrictive conditions\cite{Vio08}.
 
The WI-FIT method (``Wavelet based hIgh resolution Fitting of
Internal Templates'' \cite{Han06}) does not require a priori
templates, but takes the information about the foregrounds
by taking differences of temperature maps at different frequencies. 
However, for the application in presently available
maps, it requires the assumption that the spectral indices are
constant in space, which, as said, is a very inaccurate approximation.
Their assumption that the Galactic emission in each pixel is
proportional to the difference in temperature maps 
($(T_i^\nu - T_i^{\nu '})\propto T_i^\nu $) is in general 
incorrect because $T_i^{\nu '}$ is not proportional to $T_i^\nu $, 
the temperature in each pixel being the superposition of many different emissions
with different temperatures. Again, we have here the same problem
as with the use of templates: the assumption
that there is only one temperature along each line of sight,
and that the intensity of this emission is describable
with a simple average fixed power law multiplying a black body emission
with an average temperature. It has been claimed\cite{Han06}  that
their method is good because they obtain similar results to other authors with different 
methods\cite{Ben03b}, but this may be due to their similar
assumptions.
 
Foreground contamination residuals are found even for the best
available supposedly clean CMB maps. 
Some correlation is found 
between $\Delta \ell=4n$ and $n$=1,2
spherical harmonic multipole domain, which is caused by a symmetric 
signal in the Galactic coordinate system\cite{Nas06}.
The alignment of low-$\ell $ multipoles
appears to be rather robust to Galactic cut and different
foreground contaminations\cite{deO06}.
The statistical anisotropy in different circles of the sky
is also found\cite{The06}, pointing that 
foreground contamination residuals are found even for the best
available supposed clean CMB maps. 
Isotropy of the ILC maps is ruled out to 
confidence levels of better than 99.9\% \cite{Sam07}. 
The cross-correlation between CMBR data and $\gamma $-ray data has been studied\cite{Liu06} and it was concluded that an unknown source of radiation, most probably of galactic origin, is
implied by their analysis. This
unknown radiation of galactic origin might take place in the surface
of Galactic HI structures moving through interstellar space and/or
interacting with one another\cite{Ver07}. 
The spatial association on scales of 1--2 degrees 
between interstellar neutral hydrogen\cite{Ver07},
integrated in maps over ranges of 10 km/s and CMBR maps cleaned of foreground contamination through the ILC methods,
is especially significant for the present discussion too. Several extended
areas of excess emission at high galactic latitudes ($b>30^\circ $)
are present in both maps. These structures are thought to
have typical distances from the Sun of order 100 pc\cite{Ver07}.

Hence, considering only the Galactic dust component, the methods used to remove 
it (templates, cross-correlations, assumption of a Galactic power
spectrum, MEM, ILC, etc.) are all  inaccurate and one should
not expect to produce maps clean of Galactic dust contamination by applying
them.
The analysis of galactic latitude dependence of these anisotropies
and the fact that the power spectrum
is almost independent of the frequency over the range 50--250 GHz
can be considered at least as a proof that the Galactic dust emission
is lower than 10\% on the $\sim$1 degree scale 
[or double of this value considered to 2$\sigma $], possibly higher
for lower $\ell$ multipoles\cite{Lop07}.
This uncertainty in Galactic contamination 
may produce important systematic errors in some cosmological
parameters.  
In any case, one thing is clear: the present error bars calculated
for cosmological parameters are very significantly underestimated,
and the range of possible values is not as small as indicated by
the claims of  ``precision cosmology''.
 
Galactic contamination is also important for polarized CMBR light.
Synchrotron radiation of the Galaxy is strongly polarized; it is observed in radio (408 MHz), 
and it has loops with an excess of $\sim$20 $\mu $K, probably due to shells of very old supernova remnants\cite{Sar82}, which are not being taken into account in the subtraction of
foregrounds for polarized CMBR light\cite{Sar14}. Thermal dust also leaves an imprint on polarized light. As a matter of fact, the recent claim of the discovery of proofs of gravitational waves
due to inflation announced with great  ballyhoo, based on
the BICEP2 detection of a $\sim 0.3 \mu $K B-mode signal, was later shown to be a fake result
owing to defective analysis of Galactic contamination. The measurements were done
 in a region of a loop with important radio emission and with some contamination of dust too\cite{Sar14,Liu14}. The anomalous radiation has the spectrum of magnetic dipole radiation.
 
\subsection{Other contaminants and anomalies}
 
Apart from the cosmological and the Galactic signal,
there may be other contaminants: either from closer sources (in the
solar system or the solar neighbourhood), 
or extragalactic sources much closer than $z=1000$--1500
(as it is supposed for the cosmological origin).
The two lowest cosmologically interesting 
multipoles, $\ell$=2 and 3, are not statistically isotropic\cite{Cop06,Cop07}. 
The planes of the quadrupole and the octopole are unexpectedly aligned
\cite{Cop06,Cop07,Su08,Sta12} at 99.6\% C.L.\cite{Cop07}. This alignment can be 
regarded as a generalization of the alignment of the $\ell=2$ and 3 modes -- 
the so-called `Axis of Evil'.
Indeed, the combined quadrupole plus octopole is surprisingly 
aligned with the geometry and direction of motion of the solar system: 
the plane they define is perpendicular to the ecliptic plane and to 
the plane defined by the dipole direction, and the ecliptic plane 
carefully separates stronger from weaker extrema, running within a 
couple of degrees of the null-contour between a maximum and a minimum over 
more than 120$^\circ $ of the sky.  
 
Moreover, the angular two-point correlation 
function at scales $>60$ degrees in the regions outside the Galactic cut 
is approximately zero in all wavebands 
and is discrepant with the best fit $\Lambda $CDM inflationary model: 
99.97\% C.L. for the discrepancy\cite{Cop07,Rak07}, which points
towards a violation of statistical isotropy \cite{Cop08}.
The lack of quadrupole CMB signal is, according to some authors\cite{Liu12,Sta12}, 
a serious challenge to the standard model.
There is abnormally high (low) powers in the $\ell=6$, 12--17 modes\cite{Su08}; 
the probabilities for having the anomalous amplitudes of the $\ell=5$, 6, 17 modes are about 
0.1\%, 1\% and 1\% respectively according to the Gaussian conjecture. 
The ratio of the 
large-scale fluctuation amplitudes in the southern ecliptic hemisphere is 
high at the level 98--99\% \cite{Eri04a}, with an absence of large-scale 
power in the vicinity
of the north ecliptic pole. This asymmetry is stable with respect
to frequency and sky coverage. Also, the
first peak in the anisotropies does not have a blackbody shape
and shows a clear hemispherical anisotropy\cite{Jia10}.

However, the WMAP (Wilkinson Microwave Anisotropy Probe) team \cite{Ben11} did not find a significant detection of hemispherical or dipole power asymmetry across the sky. They also claim that the cold spots in the map
are statistically consistent with random CMB fluctuations (a claim that is contradicted by later analyses of Planck data\cite{Pla16}); that the amplitude of the quadrupole is well within the expected 95\% C.L.; that there is no anomaly in the lack of large angular CMBR power.
The WMAP team\cite{Ben11} admit, however, that there is a remarkably unexpected 
alignment of the quadrupole and octopole.
 
A solution to explain some of these anomalies is the presence of a certain amount of contamination
in relation with the solar system or its neighbourhood,
although part of the effect may be due to non-uniform 
sky coverage\cite{Chy05} too. If a hypothetic foreground 
produced by a cold spot in the Local Supercluster is subtracted 
from the CMBR data, the amplitude of the 
quadrupole is substantially increased, and the statistically 
improbable alignment of the quadrupole with the octopole is 
substantially weakened, but this does not explain the coincidence
of the alignment with the ecliptic\cite{Abr06}.
 
Anomalies of CMBR data could also be explained by synchrotron 
radiation emitted in the heliosheath\cite{Sha09a,Sha09b,Sha09c}. Radiative and 
plasma/Mag\-ne\-to\-hy\-dro\-dy\-na\-mic processes within the heliosheath could affect the propagation 
of an otherwise isotropic CMBR background through it. 
The geometric distortion of the termination 
shock might be a possible source for the quadrupole. Free--free 
thermal radiation which involves pick up ions at the termination
shock region can explain both the blackbody quadrupole and 
the non-blackbody component to the quadrupole. 
There is also a prediction for the octopole.
 
If there were no such contamination and the tension of the low quadrupole, 
the low octopole, and the alignment of the quadrupole and octopole were real, the
tension could be alleviated with a different cosmological model; for instance,
$R_h=ct$ cosmology \cite{Mel15a}.
 
Another consideration that could point to the importance of the 
non-cosmological extragalactic contamination over photons of cosmological
origin is that the simulations of the gravitational lensing of the microwave background 
by galaxy clusters at $z<1$ under any plausible Big Bang model variation 
produces far more dispersion in the angular size of the primary acoustic peaks 
than the observations allow\cite{Lie05}. 
When all the effects are taken together, it is difficult to understand how CMBR data could 
reveal no evidence whatsoever of lensing by groups and clusters. 
Cool spots in the microwave background are too uniform in size to have 
travelled from $z=1000$--1500 to us. 
There should be a spread of sizes around the average, 
with some of these cool spots noticeably larger and others noticeably smaller. 
But this dispersion of sizes is not seen in the data. 
Too many cool spots have the same size. Moreover,
the observed Sunyaev--Zel'dovich effect caused by the clusters only 
accounts for about 1/4 of the expected decrement\cite{Lie06};
although the level of Sunyaev--Zel'dovich effect is observed
as in the predictions in radio\cite{Bon06}.
Effects such as a central cooling flow in clusters, 
the abundance of hot cluster gas, large scale radial decline 
in the temperature, uncertainties in the $\beta$-model
and the role played by cluster radio sources are too weak to change the 
estimation\cite{Lie06}.
Under Big Bang premises, 
this implies that the cosmological parameters (including the Hubble constant, 
the amount of dark matter, etc.) used to predict the original, pre-lensed 
sizes of the cool and hot spots in the microwave background might be 
wrong, or that some of these cool spot structures are caused by nearby physical 
processes and are not really remnants of the creation of the Universe; 
or perhaps there is some other, unknown factor damping the effects of dispersion 
and focusing. It was speculated that the large-scale curvature of space may 
not entirely be an initial value problem related to inflation. The absence of 
gravitational lensing of the CMBR points to the possibility that even effects 
on light caused by wrinkles in the space of the late (nearby) Universe have 
been compensated for, beyond some distance scale, by a mechanism that 
maintains a flat geometry over such scales.
Or high energy electrons may 
synchrotron radiate in the intracluster magnetic field of strength 
B $<\sim$ 1$\mu$G to produce cluster microwave emissions that account for the missing Sunyaev--Zel'dovich effect flux\cite{Lie06b}.
However, one is also tempted to interpret this in terms of the importance
of extragalactic non-cosmological contribution in the microwave anisotropies.
 
Summing up, even if we accepted a cosmological origin for most of the CMBR radiation,
putatively clean maps of CMBR anisotropies are most probably not totally free from
from solar system, Galactic, or extragalactic contamination, and an accurate
way of correcting for all these contributions has still to be devised since we do not
have accurate information on the microwave emission of any of these
contributions.

\section{Nucleosynthesis and reionization}
\label{.nucleo}
 
\subsection{Light element abundances and baryon fraction}
 
In the standard model, it is claimed that $^4$He and other light elements 
(Deuterium, $^3$He, $^9$Be, $^7$Li) were created
in the primordial Universe, and the existence of these elements is used as a proof
for the necessity of a hot Universe in the past. However, there are alternatives.
The alternatives may have some caveats to explain all observations accurately,
but neither is Big Bang theory free of problems.
 
Helium could be created with several explosions such as those in
the (quasi-)steady state theory, or could be synthesized in massive objects evolving 
in the nuclear regions of galaxies\cite{Bur71}.
The existence of massive stars in the first moments of the formation of galaxies, which
in few hundred million years would produce the 24\% helium now observed, and
this would be distributed through the interstellar space by supernova
explosions. Indeed, Burbidge \& Hoyle\cite{Bur98} have argued
that a case can be made for making all the light nuclei in stars.
Population III stars are believed to contribute to the observed 
near-infrared background and heavy element pollution of the intergalactic medium,
and it could contribute to the primordial He abundance\cite{Sal03}. 
Some theoreticians object that in such a case there should be more oxygen and carbon
than observed; this could be solved if the least massive
stars had not exploded but blew off their outer layers (pure helium)\cite{Ler91}(ch. 6).
Certain rare light isotopes cannot have 
been produced in this way, but the cosmic rays generated by early
stars, colliding with the background plasma, would generate them.
Therefore, there are alternatives to the Big Bang to produce any of these
elements, although, of course, the standard model is the proposal that is
the most complete in detail up to now.

It is said that $^4$He and $^7$Li abundances are all consistent with those expected  minutes after
the Big Bang, provided that the present universe has a baryon density in the
range $0.018<\Omega _bh^2<0.022$ \cite{Sch98b}, but this statement is not totally free of
controversy. The best known abundance is that  for $^4$He, around $Y_P=0.24$, when the metallicity
tends to zero. However, there are claims of some problems with the model:
measurements of the abundance of helium of $Y_P=0.2565\pm 0.0010$(stat.)$\pm 0.0050$(syst.), higher at the 2-$\sigma $ level than the prediction of standard 
Big Bang nucleosynthesis\cite{Izo10}. Furthermore, galaxies of poor metal content, 
implying minimal stellar contamination,
show the mass fraction of $^4$He as low as 0.21 \cite{Ter92}.
Even allowing for error bars, this is far too low to match the Big Bang 
predictions. Only an ad hoc explanation of inhomogeneities in the
primordial set-up conceals the big bang nucleosynthesis with this result.
There are also stars in our own Galaxy that have a low helium abundance in 
their atmosphere (the subdwarf B stars), but it was shown\cite{Sar67} 
that these also revealed other peculiarities, such as phosphorus
and similar elements, which were associated with the chemically peculiar 
stars of population I. It was believed that these objects do not show
a normal helium abundance and were not taken into account.
Also, anomalously low helium content (lower than primordial helium) is required to fit the luminosities and temperatures of the metal-poor K dwarfs\cite{Cas07}.
 
According to
some authors, the Big Bang also fails in its predictions of the abundance of lithium.
The Li$^7$ abundance is much lower than the prediction\cite{Rya00,Coc12,Fam12}.
The Li problem remains and is indeed exacerbated; the discrepancy is 
a factor 4.2$\sigma $ (from globular cluster stars) to 
5.3$\sigma $ (from halo field stars) \cite{Cyb08}.
This observation of the ``Spite Plateau'' (the name given to the baseline in the abundance of lithium found in old stars orbiting the galactic halo) is
relevant since it is believed that such stars were formed early in the universe out of material that had not been significantly modified by other processes.
It was suggested that this discrepancy could be caused by a modification of
surface lithium abundances during the stars' lifetimes\cite{Kor06}: the lithium
is destroyed in old, metal-poor halo stars. This idea has some support in the fact the lithium in the interstellar medium of the Magellanic Clouds,
where the destruction process cannot happen, is in agreement with Big Bang nucleosynthesis predictions (4 times larger than in metal-poor halo stars)\cite{How12}.

The evolution of deuterium and the mechanisms of possible deuterium production
from the Big Bang  are still not properly understood\cite{Vid98,Pro03}.
For beryllium, excessive abundances are observed in the stars\cite{Cas97}, 
but this is justified with the argument that they have no null
metallicity and the abundance should be lower in the primordial Universe.
It is also argued that our Galaxy produces beryllium,  the accretion of matter\cite{Cas97},
or that primordial nucleosynthesis 
should be a posteriori substituted by inhomogeneous nucleosynthesis\cite{Boy89}.
 
Still, as we have seen, some open questions are discussed
about nucleosynthesis, and whether this can be substituted by alternative models remains
to be seen. The question is not as simple as the Big Bang making a unique prediction and the
observations confirming it; there is a lot of 'cooking' behind these numbers, a lot of
astrophysical processes on element production and evolution 
that are proposed ad hoc in order to fit some observations that
do not fit the a priori predictions, and even so there remain problems to explain everything. There are also some cosmological parameters to play with: the neutron decay time, the number of neutrino species,
the ratio between baryons and photons and others; although perhaps the first two are more or
less known. Curiously, a much better fit to primordial nucleosynthesis
to observations is given when the number of neutrino species is two instead of three, as
predicted by the standard model for particles\cite{Hat95}. 
 
Perhaps the most important aspect to criticize is the methodology itself for testing the primordial
nucleosynthesis\cite{Kur92}: Instead of trying to make predictions that can
be tested by observation, cosmologists take observations that can be made, such
as measurements of the abundances in Population II material, and try to
determine the primordial abundances while knowing a priori what answer is needed
to fit some particular model. 
They then take those derived abundances
and adjust the model to match more closely so that circularity is completely
guaranteed. In this way, the baryon density derived is too low to account
for the subsequent large scale structure of the universe, and an ad hoc
addition of cold, dark non-baryonic matter, cosmological constant must be introduced.
 
Less than 50\% of the baryons predicted by the $\Lambda $CDM model to exist
at low redshift have been found in some way in the galaxies or
the intergalactic medium \cite{Tak07,And10,McG10}. 
Where is the rest of it? A second problem with baryons is that galaxies
are observed to have a significantly smaller baryon fraction relative to the cosmic
average\cite{And10,McG10}. Tentative explanations of these missing baryons
are given: a large fraction of the missing baryons may 
reside in the filaments of the cosmic web\cite{Eck15}.

\subsection{Reionization epoch}
 
In Big Bang cosmology, reionization is the process that ionized the matter in the universe after the dark ages. As the majority of baryonic matter is in the form of hydrogen, reionization usually refers to the reionization of hydrogen gas. This occurred once objects started to condense in the early universe that were energetic enough to reionize neutral hydrogen. As these objects formed and radiated energy, the universe reverted from being neutral to once again being an ionized plasma. At that time, however, matter had been diffused by the expansion of the universe, and the scattering interactions of photons and electrons were much less frequent than before electron--proton recombination. Thus, a universe full of low density ionized hydrogen would remain transparent, as is the case today.
 
The presence of diffuse neutral hydrogen should produce an
absorbing trough shortward of a QSO's Lyman-alpha emission 
line---the Gunn--Peterson effect.
A hydrogen Gunn--Peterson trough was predicted to be present 
at a redshift $z\approx 6.1$ . Indeed,
a complete Gunn--Peterson trough at $z=6.28$\cite{Bec01} was discovered, 
which means that the Universe is approaching the end of the reionization
epoch at $z_r\approx 6$\cite{Bec01}. 
The quasar population is unlikely to provide enough
photons to ionize the universe at $z\approx 6$ \cite{Fan01b}, even less for
higher redshift, where there is a lower number of QSOs.
At $z\approx 6$, the amount of absorption increases quickly with
redshift (a tentative detection of a complete Gunn--Peterson trough),
indicating that $z\approx 6$ is close to the end of the reionization epoch.
However, the $z=6.5$ Lyman-alpha lines are not strongly suppressed by a neutral intergalactic medium, indicating that reionization was not largely complete at $z=6.5$ \cite{Mal04}, and galaxies have been observed at much higher redshifts without the opacity features prior to
the reionization, so the epoch of reionization was moved beyond those higher redshifts. 
An inhomogeneous reionization\cite{Bec01} is a possibility to explain
the apparent disagreement of the different data.
 
Measurements of CMBR anisotropies give a reionization
epoch $z_r= 10.5\pm 1.2$ (with WMAP data\cite{Jar11}) or 
$z_r=8.8^{+1.3}_{-1.2}$ (with Planck data\cite{Pla15b}).
This might be indicative that reionization is likely to have been extended in time;
otherwise, we would have a tension here.
 
It is thought that reionization might be due to light from massive stars.
But this is still a problem for reionization at $z\approx 10$ because
the measured star formation density is found to decrease too quickly with
increasing redshift\cite{Bun10,Bou11}.
Another exotic and ad hoc proposal to solve the problem uses, for instance, fast accretion shocks formed around the cores of the most massive 
haloes\cite{Dop11}.
 
\section{Formation of galaxies and dark matter problem}
 
\subsection{Large scale structure}
 
Like in all the other sections,
there are other theories which explain the formation of voids and
structure: radiation given off by primordial galaxies and QSOs\cite{Kur92};
galaxies that beget other galaxies\cite{Bur97b}; 
plasmas in the middle of magnetic fields and electric currents that create
filaments, and this would explain the filamentary Universe\cite{Ler91,Bat97a,Flo97,Bat97b,Bat98};
the Quasi steady State Theory\cite{Nay99}, which reproduces the observed 
two-point correlation function; and others.
 
Even if we ignore the alternative scenarios, we cannot say that everything
is well undestood in the standard one. Caveats or open questions are still present.
The standard theory would require large-scale homogeneity on scales of 
distance greater than a few tens of Mpc, and the distribution of
galaxies and clusters of galaxies should be random on large scales. However,
the departures from homogeneity claiming a fractal Universe and the regularity
of structure are frequent.
Pencil-beam surveys show large-scale structure out to distances of more 
than 1 Gpc in both of two opposite directions from us. This appears as a 
succession of wall-like galaxy features at fairly regular intervals, with 
a characteristic scale of 128 $h^{-1}$ Mpc\cite{Bro90}, the 
first of which, at about 130 Mpc distance, is called ``The Great Wall''. 
Several such evenly-spaced ``walls'' of galaxies have been found\cite{Kur90}. 
The apparent lack of periodicity in other directions led to the initial
report being regarded as a statistical anomaly\cite{Kai91}, but
a reconfirmation of a $(120\pm 15)h^{-1}$ Mpc periodicity for clusters
of galaxies came after that (e.g.\ \cite{Ein97}).
An analysis of the distribution $N(z)$ of photometric redshifts in a grid of the deep fields 
reveals the possible existence of super-large structures with a contrast $dN/N\sim 50$\% 
and tangential and radial dimensions of about 1000 Mpc \cite{Nab10,Mas07}. 
 
Some observations show large scale inhomogeneities and peculiar velocities typical of a fractal 
on scales up to at least 100 Mpc, so the density fluctuations inside the
fractal inhomogeneity cell will lead to a strong disturbance of pure Friedmann
behaviour\cite{Hag72,Fan91,Rib92a,Rib92b,Rib93}. 
Hence, the Universe is not close to the 
Friedman--Lema\^\i tre--Robertson--Walker on scales less than 100 Mpc.
Statistical differences from homogeneous 
distribution were even found out to a scale of at least 200 Mpc\cite{Bes00}. 
Structure is dominated by filaments
and voids on this level\cite{deL86} and by large velocity
flows relative to the cosmological background\cite{Dre87}.
However, other observations suggest the opposite conclusion:
a striking linearity of the Hubble law in the distance range between
2 and 25 Mpc\cite{San86,Pee98,Ekh01,Kar02}. 
Beside the paradox, parameters such $H_0$, $q_0$
and $\Omega _0$ are determined without knowing on which scales the radial
motion of galaxies and clusters of galaxies relative to us is completely
dominated by the Hubble flow\cite{Mat95}.
Most cosmologists appeal to the highly isotropic character of
the microwave background as one of the principal justifications for
assuming that the Universe is homogeneous on large scales. By itself,
the fact that some observer sees isotropic background radiation is
inconclusive, for this can be true in a static inhomogeneous universe, as well
as in a spherically symmetric inhomogeneous universe where we are near
a centre of symmetry\cite{Mat95}. So an open question
is apparently present.
 
There is streaming inconsistent with the Local Group absolute space velocity 
inferred from the CMBR dipole anisotropy\cite{Lau94}.
The standard model interprets this as the existence of a puzzling group 
flow of galaxies relative to the microwave radiation on scales of at least 
130 Mpc. Earlier, the existence of this flow led to the hypothesis of a 
``Great Attractor'' pulling all these galaxies in its direction. But in newer studies, no backside infall was found on the other side of the hypothetical feature, possibly because it is difficult to observe given that it is behind the Galactic plane. But if there were no attractor, the only alternative within the standard model to the apparent result of the large-scale streaming of galaxies would be that we are almost at rest, and that the microwave radiation is in motion relative to us. Either way, this result would be a trouble for the orthodox interpretation.
 
The local streaming motions of galaxies are too high for a finite universe 
that is supposed to be everywhere uniform.
In the early 1990s, we learned that the average redshift for galaxies of a 
given brightness differs on opposite sides of the sky\cite{Mat92,Lin92,Fin93,Hud99}. 
There is also the Bullet Cluster problem: the observed large relative velocity of the two interacting
galaxy clusters is not accountable for in the standard cosmological model\cite{Lee10,Tho12}. 
There are anomalously large bulk flows on scales of 
100 Mpc/h and beyond, which challenge the standard $\Lambda$CDM concordance 
model as well as a large class of competitive models of dark energy 
and modified gravity\cite{Aya09}.
Results from the measurement of
large-scale peculiar velocities of clusters of galaxies using 
the largest all-sky X-ray cluster catalogue combined to date and 
CMBR data\cite{Kas08b} find a strong and coherent bulk flow on 
scales out to at least $>300 h^{-1}$ Mpc. This flow is difficult 
to explain by gravitational evolution within the framework of the 
concordance $\Lambda$CDM model. The results are confirmed with recent ``Planck?' CMBR
data\cite{Atr14}, which
show that the observed dipole of kinetic Sunyaev--Zel'dovich effect 
that gives place to the dark flow
is not contaminated by thermal Sunyaev--Zel'dovich effect. This dipole was not found with SNIa 
peculiar velocities (once the Hubble flow is removed) but the sample is
quite inhomogeneous. Possible interpretations: i) an intrinsic dipole
of CMBR in the last scattering surface (but the prediction of inflation
gives a much lower dipole); ii) the flow is real. Both interpretations are problematic.
 
Another matter is the void problem: the Local Void is observed to be too empty in comparison to the
standard cosmology predictions\cite{Tik09,Pee10}. 
Standard cosmology also predicts more dwarf or irregular galaxies in voids than observed\cite{Per08}.
 
Then there is the problem with profiles of Cluster Haloes: $\Lambda$CDM predicts a shallow low 
concentration and density profiles in contrast to observations that 
indicate a denser high concentration cluster haloes\cite{Per08}.
 
Baryonic Acoustic Oscillations (BAO) found in the Large Scale Structure\cite{And14} are a good point in favour of the standard model. But the BAO peak position depends slightly on the enrironment of galaxies,
whether in superclusters or not, which is not predicted by the standard model\cite{Rou15}.
And analyses of the Alcock--Paczynski test using Baryonic Acoustic 
BAO peaks\cite{Mel15} exclude $\Lambda$CDM within 98\% C.L.

\subsection{Periodicity in large-scale structure}
 
In a homogeneous and isotropic universe we also expect the redshift distribution
of extragalactic objects to approximate to a continuous and
aperiodic distribution. However,
Tifft\cite{Tif76,Tif77,Tif80} affirmed that there was a periodicity of 
70--75 km/s in the redshift
of the galaxies. 
In an improved correction
for the optimum solar vector, the periodicity is found to be 37 km/s (which includes
Tifft's result for twice this velocity) with a
probability of finding this period by chance of $2.7\times 10^{-5}$ \cite{Gut96}.
A galactocentric periodicity of 37 km/s was confirmed to
exist in the redshift distribution
of nearby spiral galaxies\cite{Nap06}  
The periodicity
is seen in all the datasets examined. Using the distance information, 
there is some evidence
that the periodicity is not strictly with respect to a fixed galactocentric velocity vector, but is referred to a vector that varies with increasing distance. 
 
A periodicity with $\Delta z=0.031$ or 0.062
was also found for the QSOs\cite{Bur72,Bel02c} and a 0.089 periodicity in
$log_{10}(1+z)$ \cite{Nap06}. 
Other works have also found similar periodicities\cite{Bel03a,Bel03b,Bel06b,Har07a,Har07b,Ful12}.
However, other authors\cite{Haw02,Tan05,Tan08,Baj06,Rep12} find no significant periodicity.
 
An oscillating scalar field model\cite{Hir10} with
mass with particles of mass $3.2\times 10^{-31}$ eV has been proposed
to explain the the periodic structure in galaxy counts vs.\ redshift.
Also, in 1990 the Finnish physicist Ari Lehto had come up
with a general formula for the quantization of physical quantities, without knowing of Tifft's astronomical studies\cite{Bar12}(\S 9.5.2). He found
that his formula describes well the redshift periods 36 and 72 km/s that Tifft has derived. If the energy of any photon is quantized, as Lehto suggests, then so are the frequencies, hence the redshifts of light would occur at preferred values. Lehto had searched for a common rule for the properties of the micro- and the macroworld
and he found that the ratios of the quantities involving lengths or energies may be
expressed as $2^{n/3}$ (n = 1, 2, 3, . . .). What he did was to take the ratios of observed
values of various stationary discrete systems (such as Planck energy/electron rest mass
etc.) and he noted that the exponents of two thus obtained seemed to group near
0, 1/3 and 2/3 \cite{Leh90}. In a more extensive study\cite{Leh09}, involving
over 40 quantities, it was shown that it is quite unlikely that the groupings could
be due to chance. The pattern is best revealed by quantities that have been most
accurately measured (laboratory physics), but it seems to be visible also for cosmic
quantities. Lehto's system has the Planck units as natural starting points for making up the
physical world. The Planck scale is absolute, based on constants of nature. For example,
the observed temporary periods take values $t = t_{Pl}2^{n/3}$, and the electron
mass is obtained as $m_e = m_{Pl}2^{227/3}$. The redshift steps of about 72
and 36 km/s would come from $V = 2^{n/3}c$ where n/3 is 12 and 13.

\subsection{Antimatter and inflation}
 
Protons do not decay, as far as we know from particle physics experiments
until now. Therefore,
the Universe should be made up equally of matter and antimatter according to
the standard model. Matter dominates the present universe apparently 
because of some form of asymmetry, such as CP violation asymmetry, that caused 
most anti-matter to annihilate with matter, but left much residual matter. Experiments 
are searching for evidence of this asymmetry, so far without success. Other 
galaxies cannot be antimatter because that would create a matter--antimatter 
boundary with the intergalactic medium that would create gamma rays, which 
are not seen\cite{Tau97}. Another mystery of the standard model.
 
The introduction of inflation into the standard theory is another sign of the weakness of
the model. The inflation necessary to explain a flat Universe is very improbable\cite{Ste11}.
And inflation has no predictive power because anything may happen with inflation. A theory that can predict anything is a theory that predicts
nothing. There are indeed $\sim$1000 models of inflation\cite{Mart13}.

\subsection{Dark matter and the inconsistencies of the theory at 
galactic scales}
 
Zwicky\cite{Zwi33} paper in 1933 on dark matter in rich clusters proposed an application of the
virial theorem to these data and it gave 
a mass-to-light ratio of $\sim$50 in solar units.
Kahn \& Woltjer\cite{Kah59} in 1959 determined the mass of the Local Group
and obtained a mass-to-light ratio of 43 in solar units.
In the '50s, Page\cite{Pag52,Pag60} also found that pairs of elliptical galaxies had a mass-to-light
ratio of $66\pm 14$ in solar units. This showed that such binaries must have
massive envelopes or be embedded in a massive common envelope. Similar
results were obtained in the '50s from 26 binary galaxies by Holmberg\cite{Hol54}.
Babcock\cite{Bab39} first showed in 1939 the need for dark matter for an individual
galaxy, by measuring the rotation curve of the outer regions of M31.
However, the majority of astronomers did not become convinced of the need for dark
matter in galaxies until the publication of works in the '70s on
the stability of galactic discs
by Ostriker \& Peebles\cite{Ost73}, the paper in which 
Ostriker et al.\cite{Ost74}  
showed that the mass of spiral galaxies increases almost linearly with radius
to nearly 1 Mpc, and rotation curves by Vera Rubin\cite{Rub80}. Once again, this shows the typical mentality of astrophysicists:
accepting facts only when there is a theory supporting it with an explanation, a not-so-empirical approach that dominates the development of cosmology.
 
That there is some dark matter, either baryonic or non-baryonic, is clear, 
but how much, and what is its nature?
The success of the standard model in converting a hypothesis into a solid theory
depends strongly on the answer to these open questions.
Stellar and cold gas in galaxies sum to baryonic matter content that is 8$^{+4}_{-5}$\%
of the total amount of the predicted Big Bang baryonic matter\cite{Bel03}. 
Where is the rest of the baryonic material? From here
stems another open question. Baryonic matter amounts to only around a tenth 
of the total amount of matter\cite{Tur02}.
What is the nature of the putative non-baryonic dark matter required
to achieve the current value of $\Omega _m\approx 0.3$?

Current CDM models predict the existence 
of dark matter haloes for each
galaxy whose density profile falls approximately as $r^{-2}$, although the
original idea\cite{Whi78} concerning hierarchical structures with
CDM, which gave birth to the present models, was that the dark matter was
distributed without internal substructure, more like a halo with galaxies
than galaxies with a halo\cite{Bat00}, something similar to the scenario
in Refs. \cite{Lop99,Lop02b}. 
 
Some authors have been led to question the very existence 
of this dark matter on galactic scales since its evidence is
weak\cite{Bat00,McG00,Eva01,Tas02} and the predictions do not fit the
observations: 
CDM has a ``small scale crisis'' since there are some features of the galaxies that
are very different from the predictions of the cosmological model. Nonetheless, many
researchers are eagerly trying to find solutions that make data and model compatible, 
assuming a priori that the model ``must be'' correct. Some of the problems are the following.
 
There is a problem with a observed lower density of the halo in the inner
galaxy than predicted. Profiles of Galaxy Haloes: $\Lambda$CDM 
predicts halo mass profiles with cuspy cores 
and low outer density, while lensing and dynamical observations 
indicate a central core of constant density and a flattish high dark 
mass density outer profile\cite{Per08}.
The possible solutions of core-cusp 
problem without abandoning the standard model are: bar-halo friction, which
reduces the density of the halo in the inner galaxy\cite{Sel06};
haloes around galaxies may have undergone a compression by the  
stellar disc\cite{Gne04} or/and suffered from the effects of baryonic physics\cite{DiC14}. 
 
Another problem is that the predicted angular momentum is much less than the observed one.
Binney et al.\cite{Bin01} claim that the problem of an excess of predicted 
dark matter within the optical bodies and the fact that the observed 
discs are much larger than expected can be solved if a considerable mass of low angular momentum
baryons is ejected (massive galactic outflows) and the discs are formed 
later from the high angular momentum baryons which fell in the galaxy.
The conspiracy problem is also solved if the ejection begins only 
once $M_{baryons}(r)\sim M_{\rm dark\ matter}(r)$.
Another solution within the standard cosmological model 
for the angular momentum problem is the tidal interaction of objects populating the primordial voids together with the Coriolis force due to void rotation\cite{Cas15}.
 
The enclosed dynamical mass-to-light ratio increases with decreasing galaxy luminosity and surface brightness, which is not predicted by dark matter scenarios\cite{McG14}.
 
Galaxies dominate the halo with little 
substructure whereas the model predicts that galaxies should be scaled versions 
of galaxy clusters with abundant substructure\cite{DOn04,Kro10}.
Also, the distribution of satellites is 
in a plane, incompatible with $\Lambda$CDM\cite{Kro10,Kro05,Paw13}.
The current standard model of cosmology requires the Dual Dwarf
Galaxy Theorem to be true\cite{Kro12}. According to this theorem, two types of dwarf galaxies must exist: primordial dark-matter dominated (type A) dwarf galaxies, and tidal-dwarf and ram-pressure-dwarf (type B) galaxies void of dark matter. In the model, type A dwarfs are distributed approximately spherically following the shape of the host galaxy dark matter halo, while type B dwarfs are typically correlated in phase-space. Type B dwarfs must exist in any cosmological theory in which galaxies interact.
Only one type of dwarf galaxy is observed to exist: type B dwarfs lie on the baryonic
Tully--Fisher relation, which is, however, defined by the putative type A dwarfs. Young
and old type B dwarfs coincide with dE galaxies in radii and masses, and dE galaxies are
observed to be baryon-dominated although they are popularly thought to be of putative
type A. Kroupa\cite{Kro12} says that these are arguments against the 
standard model in which one cannot make the typical rebuff of incompleteness of knowledge of baryonic physics. Furthermore, there is a correlation between bulge mass and the number of luminous satellites in tidal streams\cite{Kro10,Lop16} that is not predicted by the standard model, and it is predicted by models of modified gravity without dark matter. The disc of satellites 
and bulge-satellite correlation suggest that dissipational events forming bulges 
are related to the processes forming phase-space correlated satellite populations. 
These events are well known to occur since in galaxy encounters energy and angular 
momentum are expelled in the form of tidal tails, which can fragment to form 
populations of tidal-dwarf galaxies and associated star clusters. If Local 
Group satellite galaxies are to be interpreted as Tidal Dwarf galaxies then the substructure predictions of the standard cosmological model are internally in conflict\cite{Kro10}. 
 
Perhaps, that most severe caveat to retain the hypothesis of non-baryonic cold dark matter is that, after a  long time looking for it, it has not yet been found, although non-discovery does not mean that it does not exist.
Microlensing surveys\cite{Las00} constrain the mass of the halo in our Galaxy
in the form of dim stars and brown dwarfs to be much less than that
necessary for dark matter haloes.
Some observations are inconsistent with the dominant dark matter component
being dissipationless\cite{Moo94}. Neither are black hole haloes  a consistent
scenario\cite{Moo93}. The nature of dark matter 
has been investigated\cite{Sad99} and there are no suitable candidates.
The latest attempts to search for exotic particles have  also finished without success.
For instance, it has been attempted to detect neutralinos with the MAGIC and HESS Cerenkov telescope systems for
Very High Energy Gamma Rays through their Cherenkov radiation, but so far without success 
and only 
some emission associated with the Galaxy has been found\cite{Aha06}.
Dwarf galaxies are expected to have high ratios of dark matter and
low gamma ray emission due to other astrophysical processes so the search
is focussed on these galaxies, but without positive results.
As usual, the scientists involved in these projects justify their failure
by the fact the detectors may still
be 3--4 orders of magnitude below of the possible flux of gamma rays emitted
by Dark Matter\cite{San09} and ask for more money to continue to feed their illusions.

Note also that some other dynamical problems in which dark matter has been claimed as necessary
can indeed be solved without dark matter:
galactic stability\cite{Too81}, warp creation\cite{Lop02b}, rotation 
curves\cite{Bat00,San02}. Velocities in galaxy pairs and satellites 
might measure the mass of the intergalactic 
medium filling the space between the members of the 
pairs\cite{Lop99,Lop02b} rather than the mass of dark haloes
associated with the galaxies. 
Also the dark matter necessary to solve many problems may be baryonic:
positively charged, baryonic (protons and helium nuclei) particles\cite{Dre05}, which 
are massive and weakly interacting, but only when moving at relativistic velocities; 
simple composite systems that include nucleons but 
are still bound together by comparable electric and magnetic forces\cite{May12}, making up a three-body system ``tresinos'' or four body system ``quatrinos''; antiparticles which have negative gravitational charge\cite{Haj14}, etc.

\subsection{Dark energy and the cosmological constant or quintessence}
\label{.cosmconst}

The question of the cosmological constant\cite{Pad03}, Einstein's biggest blunder considered
now to be not such a blunder, is very amusing.
Two decades ago, most cosmologists did not favour the scenarios dominated by
the cosmological constant\cite{Fuk91}.
In the eighties, the cosmological constant was many times disregarded as an 
unnecessary encumbrance, or its value was set at zero\cite{Lon86},
and all the observations gave a null or almost null value.
However, since other problems in cosmology have risen, many cosmologists
at
the beginning of the '90s realized that an $\Omega _\Lambda =0.70-0.80$ could
solve many problems in CDM cosmology\cite{Efs90}. Years later, evidence for such
a value of the cosmological constant began to arrive. A brilliant prediction
or a prejudice which conditions the actual measurements? Another open question.
 
One measurement of the cosmological constant comes nowadays from
supernovae, 
whose fainter-than-expected luminosity in distant galaxies can be explained
with the introduction of the cosmological
constant. It was criticized as being due possibly to
intergalactic dust\cite{Agu00,Goo02,Mil15}. The presence of grey dust is
not necessarily inconsistent with the measure of a supernova at $z=1.7$
(SN 1997ff)\cite{Goo02}.
Dimming by dust along the line of sight, predominantly in the
host galaxy of the SN explosion, is one of the main
sources of systematic uncertainties\cite{Kno03}.
Also, there was an underestimate of the effects of host galaxy extinction: a 
factor which may contribute to apparent faintness of 
high-$z$ supernovae is the evolution of the host galaxy extinction with $z$ \cite{Row02}; 
therefore, with a consistent treatment of host galaxy
extinction and the elimination of supernovae not observed before maximum, 
the evidence for a positive $\Lambda $ is not very significant.
Fitting the corrected luminosity distances (corrected for internal extinctions)
with cosmological models Balazs et al.\cite{Bal06} concluded that 
the SNIa data alone did not exclude the possibility of the $\Lambda=0$ 
solution.
 
Type Ia Supernovae also possibly have a metallicity dependence and
this would imply that the evidence for a non-zero cosmological constant 
from the SNIa Hubble Diagram may be subject to corrections for metallicity
that are as big as the effects of cosmology\cite{Sha01}. 
The old supernovae might be intrinsically fainter than the local ones,
and the cosmological constant would not be needed\cite{Dom00}.
As a matter of fact, some cases, 
such as SNLS-03D3bb, have an exceptionally high luminosity\cite{How06}. 
Claims have been made about the possible existence of 
two classes of Normal-Bright SNe Ia\cite{Qui07}.
If there is a systematic evolution in the metallicity of SN Ia progenitors, 
this could affect the determination of cosmological parameters. 
This metallicity effect could be 
substantially larger than has been estimated previously and could
quantitatively evaluate the importance of metallicity evolution 
for determining cosmological parameters\cite{Pod06}. 
In principle, a moderate and plausible amount of metallicity 
evolution could mimic a $\Lambda$-dominated, a flat Universe in an open, 
$\Lambda $-free Universe. However, the effect of metallicity evolution 
appears not to be large enough to explain the high-$z$ SNIa data in a 
flat Universe, for which there is strong independent evidence, 
without a cosmological constant. 
 
Furthermore, our limited knowledge of the SN properties in the U-band
has been identified as another main source of uncertainty in the
determination of cosmological parameters\cite{Kno03}.
And the standard technique with SNe Ia consists in using spectroscopic
templates, built by averaging spectra of well observed (mostly nearby)
SNe Ia. Thus, the uncertainty in K-corrections depends primarily
on the spectroscopic diversity of SNe Ia.
 
Even if we accept the present-day SN Ia analyses as correct and without any
bias or selection effect,
other cosmologies may
explain the apparent cosmic acceleration of SNe Ia without introducing a cosmological 
constant into the standard Einstein field equation, thus negating the necessity for 
the existence of dark energy\cite{Shu10}. There are four distinguishing features of these 
models: 1) the speed of light and the gravitational ``constant'' are not constant, 
but vary with the evolution of the universe, 2) time has no beginning and no 
end, 3) the spatial section of the universe is a 3-sphere, and 4) the universe 
experiences phases of both acceleration and deceleration. 
An inhomogeneous isotropic universe described by a
Lemaitre--Tolman--Bondi solution of Einstein's fields equations can also provide
a positive acceleration of the expansion without dark energy\cite{Rom07}.
Quasi-Steady-State theory predicts a decelerating 
universe at the present era, it explains successfully the 
recent SNe Ia observations\cite{Vis07}.
Carmeli's cosmology fits data for an accelerating and decelerating universe 
without dark matter or dark energy\cite{Oli06}.
Thompson\cite{Tho12b} used available measurement for the constrainst on the variation
the proton to mass electron with redshift, and with $\frac{\Delta \alpha }{\alpha }
=7\times 10^{-6}$ he finds that almost all of the dark energy models 
using the commonly expected values or parameters are excluded.
A  static Universe can also fit the supernovae data without dark energy\cite{Sor09,Ler09,Lop10a,Far10,Mar13}.
 
There are other sources of $\Omega _\Lambda $ measurement: the anisotropies
of the microwave background ra\-dia\-tion 
(see my doubts on the accuracy of using them to determine cosmological parameters in
\S \ref{.Galcont}), for instance.
In the last two decades, a lot of proofs have been presented to the community
to convince us that the definitive cosmology has $\Omega _\Lambda \approx 0.7$,
which is surprising taking into account that in the rest of the history
of the observational cosmology proofs have been presented for $\Omega _\Lambda \approx
0$. Furthermore, recent tests indicate that other values are available in the
literature. For instance, from the test angular size vs.\ redshift for
ultracompact radio sources, it is obtained that $\Lambda $ has the opposite sign\cite{Jac97}. Using the brightest galaxies in clusters, with two
independent samples,  the fit in the
Hubble diagram is compatible with a Universe without acceleration instead of $\Omega _\Lambda =0.7$ \cite{And06}.

Moreover, the actual values of $\Omega _\Lambda $ have some consistency problem 
in the standard scenario of the inflationary Big Bang. 
The cosmological constant predicted by quantum field theory has
a value much larger than those derived from observational cosmology.
This is because the vacuum energy in quantum field
theory takes the form of the cosmological constant in Einstein's equations.
If inflation took place at the GUT
epoch, the present value would be too low by a factor $\sim 10^{-108}$, and
if the inflation took place at the quantum gravity epoch, the above factor would be 
lower still at $\sim$$10^{-120}$ \cite{Wei89}.
 
The standard model has some surprising coincidences.
There is the coincidence that now the deceleration of the Hubble flow
is compensated by the acceleration of the dark energy\cite{Unz10}(Ch. 3).
The average acceleration throughout the history of the Universe is almost null\cite{Mel12}.
Another apparent paradox is that the Big Bang Model assumes that the
cosmic fluid is not only continuous but also homogeneous and isotropic, intrinsically corresponding to zero pressure and hence zero temperature. Therefore, the ideal Big Bang Model cannot describe the physical universe as having pressure, temperature, and radiation. Consequently, the physical universe may comprise matter distributed in discrete non-continuous lumpy fashion (as observed) rather than in the form of a homogeneous continuous fluid. The intrinsic absence of pressure in the ``Big Bang Model'' also
rules out the concept of ``Dark Energy'', according to some opinions\cite{Mit11}.
 
Again, everything is far from being properly understood and with  
well-constrained parameters.

\subsection{Distant galaxies, evolution, and the age of galaxies}
\label{.distgal}
 
The Big Bang requires that stars, QSOs, and galaxies in the early universe 
be ``primitive'', meaning mostly metal-free, because it requires many generations 
of supernovae to build up the metal content in stars. But the evidence 
shows the existence of even higher than solar metallicities 
in the ``earliest'' QSOs and galaxies\cite{Bec01,Fan01b,Con02}. 
The iron-to-magnesium ratio increases at higher redshifts\cite{Iwa02}.
And what is even more amazing: there is no evolution of some line ratios, including iron 
abundance\cite{Die03,Fre03,Mai03,Bar03}, between $z=6.5$ and $z=0$.
There is no evolution of CaII absorbers with redshift \cite{Sard14}(Fig. 9).
The amount of dust in high redshift galaxies and QSOs is also much higher than expected\cite{Dun03}.
 
From the comparison with a synthesis model, colours of some very red galaxies at $z>2.5$
could be related to ages of a stellar population in passively evolving galaxies older than 1 Gyr on average\cite{Cas12}. They turn out to be massive galaxies (stellar masses of $\sim 10^{11}$ M$_\odot $) that where formed on average when the Universe was very young ($<1$ Gyr), assuming the standard cosmological parameters. 
Some early-type massive extremely red objects with $z\sim 1.4$
\cite{Lon05,Tru06} are also very old: a 
galaxy with $z=1.22\pm 0.05$  has an average stellar population age of $5.0\pm 0.1$ Gyr \cite{Lon05}, 
again in the limit of the age of the
Universe at this redshift ($5.3\pm 0.2$ Gyr).
It is clear from this information that the formation of very massive elliptical
galaxies should take place at very high redshifts.
More distant ($z>2$) red galaxies have been analysed\cite{Lab05}
with observations that provide rest-frame UV-to-NIR photometry 
(derived from 0.3-8.0 $\mu $m photometry) and
it has been found that three of the 14 red galaxies were indeed old galaxies
with Single Stellar Population best-fit ages 
2.6 Gyr at $z=2.7\pm 0.4$, 3.5 Gyr at $z=2.3\pm 0.3$,
and 3.5 Gyr at $z=2.3\pm 0.3$ respectively, representative of
the average age of the stellar populations, so the oldest stars
must be still older. 
The age of the Universe in the standard model at both redshifts of $2.7\pm 0.4$
and $2.3\pm 0.3$ is $2.5\pm 0.5$ Gyr and $3.0\pm 0.4$ Gyr respectively.
We are again with the case of galaxies as old as the Universe.
Toft et al.\cite{Tof05} fitted synthesis models to the spectra of red compact
galaxies with ages 5.5, 3.5 and 1.7 Gyr for redshifts 1.2, 1.9 and 3.4 (ages
of the Universe respectively: 5.4, 3.6, 2.0 Gyr), in the limit.
Galaxies at very high redshift ($z\sim 4$)
have been analysed\cite{Rod07} to find, with optical, near-infrared 
and mid-infrared surveys, an important ratio of old 
($\sim 1$ Gyr or older) and massive ($\sim 10^{11}$ M$_\odot $) galaxies.
With similar techniques
evidence has been presented for 11 massive and evolved (0.2-1.0 Gyr)
galaxies at redshifts $4.9\le z \le 6.5$ (photom.\ redshifts)\cite{Wik08}.
This is what has been called the impossible early galaxy problem, according to which 
observations find several orders of magnitude more very massive haloes at very high
redshift than predicted, implying that these massive galaxies formed impossibly
early\cite{Ste16}.
These results are at odds with semianalytical $\Lambda $CDM models, which claim that very massive galaxies were formed much later\cite{Guo11},
although along the line suggested by the ``downsizing'' scenario of galaxy formation.
 
In view of all this evidence and the tensions within the results expected a priori, 
orthodox cosmologists have claimed that star formation began very 
early and produced metals up to the solar abundance quickly, in a few hundred Myr, and 
in which most massive galaxies assembled substantial amounts of their stellar content rapidly (in 1--2 Gyr) beyond $z\sim 3$ in very intense star formation episodes\cite{Pere08}. 
This idea may find some support in observations such as the discovery of
 a massive starburst galaxy at $z=6.34$ \cite{Rie13}, with
mass around $3\times 10^{11}$ solar masses. 
A ``maximum starburst'' converts the gas into stars 
at a rate more than 2000 times that of the Milky Way, 
a rate among the highest observed at any epoch. All well and good, but this scenario is
at odds with the predictions of the galaxy formation in hierarchical scenarios
of $\Lambda$CDM cosmological model, so the hypothesis which served to explain 
the large scale structure succesfully is failing here.
 
\section{Conclusions}
 
There are many tenets of the standard cosmological theory that are being cast into doubt in a wide-ranging literature. Certainly, many of the references that I have cited in this review may contain wrong or uncertain arguments and results,
and further research is needed to confirm or refute their statements, but it is good that we do not forget them and continue to think about these controversial topics.
 
Note that I am not defending any specific idea of the cosmos here: neither the correctness nor the wrongness of Big Bang. I am just presenting a number of sceptical 
arguments expressing certain doubts on the validity of the standard cosmology that are found in a wide literature.
We must also admit that alternative theories are not at present as competitive as the standard model in cosmology in terms of giving better explanations. 
If they were more developed, there is a possibility that they might compete in some aspects with the Big Bang theory, but efforts are made in the present-day scientific community to avoid their development\cite{Lop14b}. 
In any case, one advantage of alternative theories is that their study may lead to new useful cosmological tests, even if the theories themselves were not viable, for instance the redshift-angular size test in connection with his studies of the Steady State theory proposed by Hoyle\cite{Hoy59}. We need unquestionable facts, free from interpretation (as Lema\^itre or Hubble did when they discovered the expansion) and good physicists to propose physical theories (Einstein, Gamow, etc.). There are so many works on cosmology, that it is more and more difficult to propose a theory without discussing all previous studies. 
This paper may be of interest in this sense.
My hope is that the reader of this review may find it useful and inspiring
in order to carry out new and challenging research, rather than the usual
boring and conformist attitude in present-day cosmology. Or can you say, after reading this paper, that everything in the standard model is clear and well established and the only remaining mission is to refine the measurement of certain parameters?

\begin{acknowledgements}
Thanks are given to Fulvio Melia and the two anonymous referees for comments on a draft of this paper that helped to improve it. Thanks are given to Terence J. Mahoney for proof-reading of the text.
\end{acknowledgements}

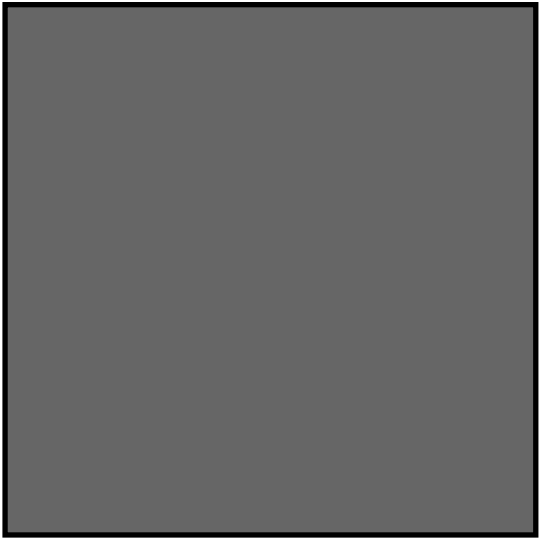
\end{document}